\documentclass[
prd 
,showpacs ,amssymb,superscriptaddress,aps
]{revtex4}
\usepackage{graphicx}
\input{epsf}

\usepackage{amsmath,amssymb}
\usepackage{bm}
\usepackage{times}

\newcommand{\dalm}{\kern1pt\vbox{\hrule height 0.9pt\hbox{\vrule width 0.9pt
\hskip 2.5pt\vbox{\vskip 5.5pt}\hskip 3pt\vrule width 0.3pt}\hrule height 0.3pt}
\kern1pt}

\newcommand{\be}{\begin{eqnarray}}
\newcommand{\ee}{\end{eqnarray}}
\newcommand{\beq}{\begin{eqnarray}}
\newcommand{\eeq}{\end{eqnarray}}

\begin{document}



\title{Strong gravitational lensing by an electrically charged black hole in Eddington-inspired Born-Infeld gravity}

\author{Hajime Sotani}
\email{sotani@yukawa.kyoto-u.ac.jp}
\affiliation{Division of Theoretical Astronomy, National Astronomical Observatory of Japan, 2-21-1 Osawa, Mitaka, Tokyo 181-8588, Japan}

\author{Umpei Miyamoto}
\affiliation{Research and Education Center for Comprehensive Science, Akita Prefectural University, Akita 015-0055, Japan}

\date{\today}

\begin{abstract}
We systematically examine the properties of null geodesics around an electrically charged, asymptotically flat black hole in Eddington-inspired Born-Infeld gravity, varying the electric charge of black hole and the coupling constant in the theory. We find that the radius of the unstable circular orbit for massless particle decreases with the coupling constant, if the value of the electrical charge is fixed. Additionally, we consider the strong gravitational lensing around such a black hole. We show that the deflection angle, the position angle of the relativistic images, and the magnification due to the light bending in strong gravitational field are quite sensitive to the parameters determining the black hole solution. Thus, through the accurate observations associated with the strong gravitational lensing, it might be possible to reveal the gravitational theory in a strong field regime.  
\end{abstract}

\pacs{04.50.Kd, 04.40.Nr, 04.70.-s}
%
\maketitle
\section{Introduction}
\label{sec:I}

Gravity is a fundamental interaction, and there are many observations and experiments to test the gravitational theories, some of which seem to rule out theories alternative to the general relativity. 
These tests did not show any defects in general relativity, proposed by Einstein, while they have been implemented basically in a weak field regime \cite{W1993}. On the other hand, unlike the tests in a weak field regime, the observations in a strong field regime are quite restricted, which are still not enough to verify the gravitational theory in such a strong field regime. That is, the gravitational theory might be modified in the strong field regime, which can lead to the phenomena depending on the detail of the gravitational theories. If so, one will be able to probe the gravitational theory via the observations associated with compact objects. In practice, there are some suggestions for distinguishing the gravitational theories by using astronomical observations (e.g., \cite{P2008,SK2004,S2009,YYT2012,S2014}).

So far, many modified theories of gravity have been proposed, where some of them are formulated in order to solve open problems in general relativity. Eddington-inspired Born-Infeld theory of gravity (EiBI) is also one of them \cite{Banados:2010ix}, with which one can avoid not only the big bang singularity \cite{AF2012,BFS12013}, but also the singularity appearing due to the gravitational collapse of dust \cite{PCD2011}. EiBI is based on the Eddington gravity \cite{E1924} and on the nonlinear electrodynamics of Born and Infeld \cite{BI}. In EiBI, the connection is considered as an independent field as well as the metric, following a Palatini formalism. One of the interesting features of EiBI is that the theory in vacuum completely agrees with general relativity. One can observe the deviation of EiBI from general relativity with matter, where such deviation becomes significant especially inside neutron stars. That is, one could distinguish EiBI from general relativity via the observations of neutron stars itself and/or the phenomena associated with neutron stars \cite{PCD2011,PDC2012,SLL2013,HLMS2013,Sotani:2014goa,SLL2012,Sotani:2014xoa,S2015}.

The discussions about the neutron stars in EiBI are lively since EiBI was proposed, while the examinations with respect to the black holes in EiBI are limited. This might be because the simplest black hole solution in EiBI, i.e., the spherically symmetric spacetime in vacuum, agrees with the Schwarzschild black hole in general relativity. Nonetheless, EiBI with non-zero energy-momentum tensor can deviate from general relativity even if there is no matter. In other words, en electrically charged black hole in EiBI is expected to be different from that in general relativity. Such a black hole solution is discussed in Refs. \cite{Banados:2010ix,Wei:2014dka,SM2014,FL2015}.

With respect to the phenomena around black holes, the light bending due to strong gravitational field is one of the important properties from the observational point of view. In particular, when the light passes in the vicinity of the black hole, the deflection angle can become more than $2\pi$, where the light goes around the black hole before approaching the observer. These phenomena are known as the strong gravitational lensing. In general relativity, the properties of such phenomena are examined well around various black hole solutions \cite{VE2000,ERT2002,B2002}. In a similar vein, recently the strong gravitational lensing around the electrically charged black hole in EiBI is partially examined \cite{Wei:2014dka,JK2015}. However, such examinations might be insufficient, where only the case of the positive coupling constant in EiBI is considered \cite{Wei:2014dka} or the angular position and magnification of the relativistic images due to the gravitational lensing are not determined \cite{JK2015}. Thus, in this paper, we systematically examine the properties of strong gravitational lensing around the electrically charged black hole in EiBI, varying the both positive and negative coupling constant in EiBI in addition to the value of the electrically charge. For this purpose, we derive the analytic formulae describing the strong gravitational lensing, basically according to Ref. \cite{B2002}. Then, we will show that the various properties in EiBI are significantly sensitive to not only the coupling constant in EiBI but also the electrically charge of black hole. In this paper, we adopt geometric units, $c=G=1$, where $c$ and $G$ denote the speed of light and the gravitational constant, respectively, and the metric signature is $(-,+,+,+)$.

\section{Electrically Charged Black Hole in EiBI}
\label{sec:II}

EiBI proposed by Ba\~nados and Ferreira \cite{Banados:2010ix} is described with the action
\begin{equation}
  S = \frac{1}{8\pi\kappa}\int d^4x \left(\sqrt{|g_{\mu\nu} + \kappa R_{\mu\nu}|} - \lambda\sqrt{-g}\right) + S_{\rm M}[g,\Psi_{\rm M}], \label{eq:action}
\end{equation}
where $R_{\mu\nu}$ is the symmetric part of the Ricci tensor constructed by the connection $\Gamma^\mu_{\alpha\beta}$, $|g_{\mu\nu} + \kappa R_{\mu\nu}|$ denotes the absolute value of the determinant of the matrix $(g_{\mu\nu} + \kappa R_{\mu\nu})$, $g$ denotes the determinant of the metric $g_{\mu\nu}$, $S_{\rm M}$ denotes the action of matter fields, and $\Psi_{\rm M}$ collectively denotes any matter field. As seen in Eq. (\ref{eq:action}), EiBI has two parameters, i.e., $\kappa$ and $\lambda$. $\kappa$ is the Eddington parameter with the dimension of length squared, while $\lambda$ is a dimensionless constant associated with the cosmological constant $\Lambda$ as $\Lambda=(\lambda-1)/\kappa$. Additionally, since the action of EiBI in the limit of $\kappa=0$ or for $S_{\rm M}=0$ reduces to the Einstein-Hilbert action \cite{Banados:2010ix,Pani:2012qd,Harko:2014oua}, EiBI in the limit of $\kappa=0$ and/or $S_{\rm M}=0$ becomes equivalent to general relativity with the cosmological constant. The Eddington parameter $\kappa$ is constrained via the solar observations, big bang nucleosynthesis, and existence of neutron stars \cite{Banados:2010ix,Pani:2011mg,Casanellas:2011kf,Avelino:2012ge}. In addition to these constraints, the possibilities to observationally constrain $\kappa$ are suggested with the simultaneous measurements of the stellar radius of the $0.5M_\odot$ neutron star and the neutron skin thickness of ${}^{208}$Pb \cite{Sotani:2014goa}, and with the frequencies of the neutron star oscillations \cite{Sotani:2014xoa}.

Since in EiBI the connection $\Gamma^\mu_{\alpha\beta}$ is considered as a field independent of the metric $g_{\mu\nu}$, the field equations can be obtained from the variations of the action (\ref{eq:action}) with respect to the connection and metric \cite{Banados:2010ix};
\begin{gather}
  q_{\mu\nu} = g_{\mu\nu} + \kappa R_{\mu\nu}, \label{eq:2} \\
  \sqrt{-q}q^{\mu\nu} = \lambda\sqrt{-g}g^{\mu\nu} - 8\pi\kappa\sqrt{-g}T^{\mu\nu}, \label{eq:3}
\end{gather}
where $q_{\mu\nu}$ is an auxiliary metric associated with the connection as $\Gamma^\mu_{\alpha\beta} = q^{\mu\sigma}\left(q_{\sigma\alpha,\beta} + q_{\sigma\beta,\alpha} - q_{\alpha\beta,\sigma}\right)/2$, $q$ is the determinant of $q_{\mu\nu}$, and $T^{\mu\nu}$ denotes the standard energy-momentum tensor. We remark that $q^{\mu\nu}$ is just matrix inverse of $q_{\mu\nu}$, i.e., $q^{\mu\nu}\neq g^{\mu\alpha}g^{\nu\beta}q_{\alpha\beta}$ and $q^{\mu\alpha}q_{\nu\alpha}=\delta^{\mu}_{\nu}$. Meanwhile, raising and lowering indices in $T_{\mu\nu}$ should be done with the physical metric $g_{\mu\nu}$.

The metric describing the spherically symmetric objects is expressed as
\begin{gather}
  g_{\mu\nu}dx^\mu dx^\nu = -\psi^2 fdt^2 + f^{-1}dr^2 + r^2d\Omega^2, \label{eq:g} \\
  q_{\mu\nu}dx^\mu dx^\nu = -G^2Fdt^2 + F^{-1}dr^2 + H^2d\Omega^2,
\end{gather}
where $\psi$, $f$, $G$, $F$, and $H$ are functions of $r$, and $d\Omega^2\equiv d\theta^2 + \sin^2\theta d\phi^2$. To consider an electrically charged black hole, we adopt the energy-momentum tensor given by $T_{\mu\nu} = \left(g^{\alpha\beta}F_{\mu\alpha}F_{\nu\beta} - g_{\mu\nu}F_{\alpha\beta}F^{\alpha\beta}/4\right)/4\pi$. Then, one can derive the electrically charged black hole solution in EiBI as
\begin{gather}
  f = -\frac{r\sqrt{\lambda r^4 + \kappa Q^2}}{\lambda r^4 - \kappa Q^2} \left[\int \frac{(\Lambda r^4 - r^2 + Q^2)(\lambda r^4 - \kappa Q^2)}{r^4\sqrt{\lambda r^4 + \kappa Q^2}}dr + 2\sqrt{\lambda}M \right], \label{eq:fr1} \\
  \psi = \frac{\sqrt{\lambda} r^2}{\sqrt{\lambda r^4 + \kappa Q^2}}, \\
    E_\mu = \left(0,\frac{Q}{r^2\sqrt{f}},0,0\right),
\end{gather}
where $M$ and $Q$ denote the mass and electric charge of the black hole, while $E_{\mu}$ is the electric field outside the black hole \cite{Banados:2010ix,Wei:2014dka,SM2014}. With these metric functions in $g_{\mu\nu}$, the metric functions in $q_{\mu\nu}$ are determined by
\begin{gather}
  F = f\left(\lambda - \frac{\kappa Q^2}{r^4}\right)^{-1}, \label{eq:Fr} \\
  G = \psi\left(\lambda - \frac{\kappa Q^2}{r^4}\right), \label{eq:Gr} \\
  H = r\sqrt{\lambda + \frac{\kappa Q^2}{r^4}}. \label{eq:Hr}
\end{gather}
We remark that the black hole solution in the limit of $Q=0$ reduces to the Schwarzschild-(anti-) de Sitter spacetime in general relativity, i.e.,
$f(r)=1-2M/r-\Lambda r^2/3$ and $\psi(r)=1$, while in the limit of $\kappa=0$ reduces to the Reissner-Nordstr\"{o}m-(anti-)de Sitter spacetime in general relativity, i.e., $f(r)=1-2M/r+Q^2/r^2-\Lambda r^2/3$ and $\psi(r)=1$. Hereafter, for simplicity, we consider the asymptotically flat black hole solution, i.e., $\lambda=1$ ($\Lambda=0$). In this case, there is a maximum value of $Q/M$ depending on $\kappa/M^2$ below which the black hole solution can exist. Such maximum values of $Q/M$ are shown as a function of $\kappa/M^2$ in Fig. 5 in Ref. \cite{SM2014}.

\section{Null geodesic}
\label{sec:a2}

We consider the geodesic equation around the electrically charged black hole in EiBI. Since the Lagrangian given by ${\cal L}=g_{\mu\nu}(dx^\mu/d\tau)(dx^\nu/d\tau)/2$ is conserved along the geodesic, i.e., $ d{\cal L}/d\tau = 0 $, one can put ${\cal L} = -1/2$ for a massive particle using the scaling degree of freedom of $\tau$ \cite{SM2014}, while ${\cal L} = 0$ for a massless particle independently of the scaling degree of freedom of $\tau$. The motion of particle is subject to the Euler-Lagrange equation,
\begin{equation}
  \frac{\partial {\cal L}}{\partial x^\mu} -\frac{d}{d\tau}\left(\frac{\partial {\cal L}}{\partial \dot{x}^\mu}\right) =0. \label{eq:EL}
\end{equation}
On the other hand, adopting the metric ansatz (\ref{eq:g}), we obtain
\begin{equation}
  2{\cal L} = -\psi^2f\dot{t}^2 + f^{-1}\dot{r}^2 + r^2(\dot{\theta}^2+\sin^2\theta\dot{\phi}^2).
\end{equation}
Due to the nature of the static spherically symmetric spacetime, $\partial {\cal L}/\partial t = \partial {\cal L}/\partial \phi = 0$ in Eq. (\ref{eq:EL}), which leads to that the derivatives of $t$- and $\phi$-components of the four velocity $p_\mu$ with respect to $\tau$ become zero, i.e., $\dot{p}_t=\dot{p}_\phi=0$, where $p_\mu\equiv \partial {\cal L}/\partial \dot{x}^\mu=g_{\mu\nu}\dot{x}^\nu$. Thus, one can get 
\begin{equation}
  \dot{t}=\frac{e}{f\psi^2}\ \  {\rm and}\ \ \dot{\phi}=\frac{\ell}{r^2\sin^2\theta}, \label{eq:tphi}
\end{equation}
where $e$ and $\ell$ are constants corresponding to the energy and angular momentum, respectively, of the massless particle. In the case of a massless particle  (${\cal L}=0$), rescaling the affine parameter as $\tau \to \tau/e$ and introducing the impact parameter $b\equiv \ell / e$, the equation for the radial motion becomes
\be
	\psi^2\left( \frac{dr}{d\tau} \right)^2 + V = 1,
\;\;\;
	V(r) = \frac{ b^2 f\psi^2 }{r^2}, \label{eq:geodesic}
\ee
where to derive Eq. (\ref{eq:geodesic}) we assume that the motion is confined on the equatorial plane ($\theta=\pi/2$). We remark that the orbit of massless particle depends only on the value of $b$ independently of the individual values of $e$ and $\ell$.

The unstable circular orbit (UCO) or the photosphere, whose radius is denoted by $R_{\rm UCO}$, is obtained by solving $dV/dr = 0$ for  $d^2V/dr^2<0$. Then, the massless particle moves on the UCO around the black hole with infinite time, if $V(R_{\rm UCO}) = 1$.
Otherwise, the particle infalling from the infinity either plunges into the black hole if $V(R_{\rm UCO}) < 1$ or is scattered off by the potential barrier at some radius larger than $R_{\rm UCO}$ if $V(R_{\rm UCO}) > 1$. The condition $V(R_{\rm UCO}) \gtreqless 1 $ becomes equivalent to $ b^2 \gtreqless b_c^2 $, where $b_c$ is a critical impact parameter depending merely on the black hole parameters. We remark that, for the  Reissner-Nordstr\"om case ($\kappa=0$), $R_{\rm UCO}$ and $b_c$ can be obtained analytically as
\begin{gather}
	R_{\rm UCO} = \frac{3M}{2} \left( 1+\sqrt{ 1- \frac{8Q^2}{9M^2} } \right), \\
	b_c^2 = \frac{ R_{\rm UCO}^4 }{ ( R_{\rm UCO} -r_+ ) ( R_{\rm UCO} -r_- ) },
\end{gather}
where $ r_\pm = M \pm \sqrt{M^2-Q^2} $.

\begin{figure}
\begin{center}
\includegraphics[scale=0.5]{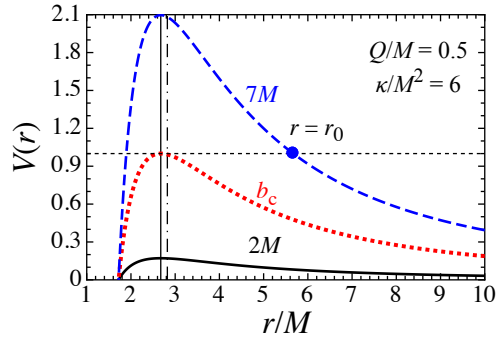} 
\end{center}
\caption{
Effective potential $V(r)$ in EiBI with $Q/M=0.5$ and $\kappa/M^2=6$. The solid, dotted, and broken lines correspond to the cases for $b=2M$, $b_c$, and $7M$, respectively. The position of $R_{\rm UCO}$ in EiBI is denoted by the vertical solid line, while that in general relativity is denoted by the vertical dot-dashed line for reference. The massless particle with $b=7M$ is scattered by the black hole at $r=r_0$, which is denoted by the filled circle.
}
\label{fig:Vr}
\end{figure}

As an example, in Fig. \ref{fig:Vr}, we show the effective potential $V(r)$ in EiBI with $Q/M=0.5$ and $\kappa/M^2=6$, where the solid, dotted, and broken lines correspond to the cases for $b=2M$, $b_c$, and $7M$, respectively. In the same figure, the position of $R_{\rm UCO}$ in EiBI is denoted by with the vertical solid line, where the position of $R_{\rm UCO}$ in general relativity, i.e., for $Q/M=0.5$ and $\kappa=0$, also denoted by the vertical dot-dashed line for reference. From this figure, one can see that $R_{\rm UCO}$ in EiBI can deviate from that in general relativity. In fact, $R_{\rm UCO}=2.67M$ in EiBI with $Q/M=0.5$ and $\kappa/M^2=6$, while $R_{\rm UCO}=2.82M$ in general relativity, i.e., the deviation is $5.3\%$. The value of $b_c$ in EiBI also can deviate from that in general relativity, i.e., $b_c=4.83M$ in EiBI  with $Q/M=0.5$ and $\kappa/M^2=6$ while $b_c=4.97M$ in general relativity. Additionally, the particle with $b>b_c$ is scattered by the black hole, as mentioned the above, i.e., for instance, the particle with $b=7M$ is scattered at $r=r_0$, which is also denoted by the filled circle in Fig. \ref{fig:Vr}.

\begin{figure*}
\begin{center}
\begin{tabular}{cc}
\includegraphics[scale=0.5]{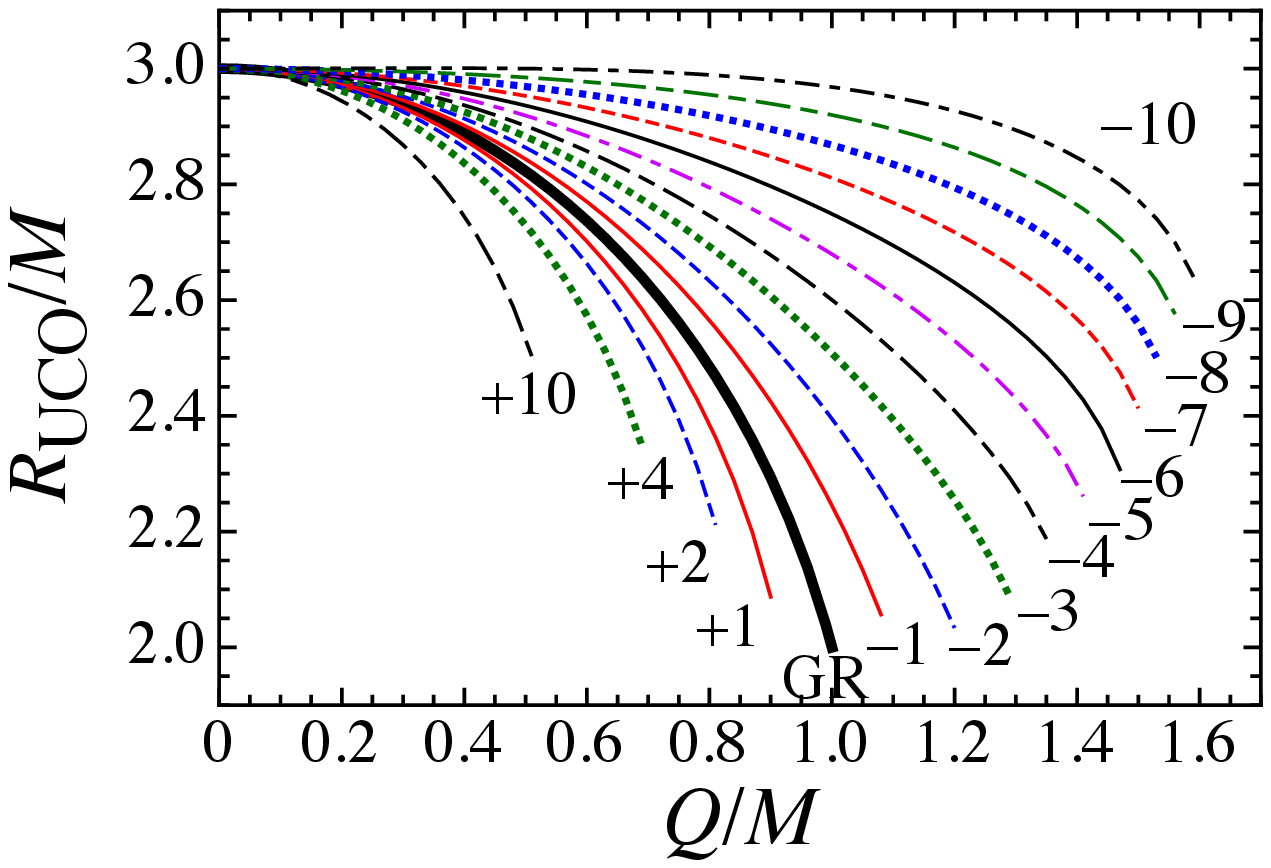} &
\includegraphics[scale=0.5]{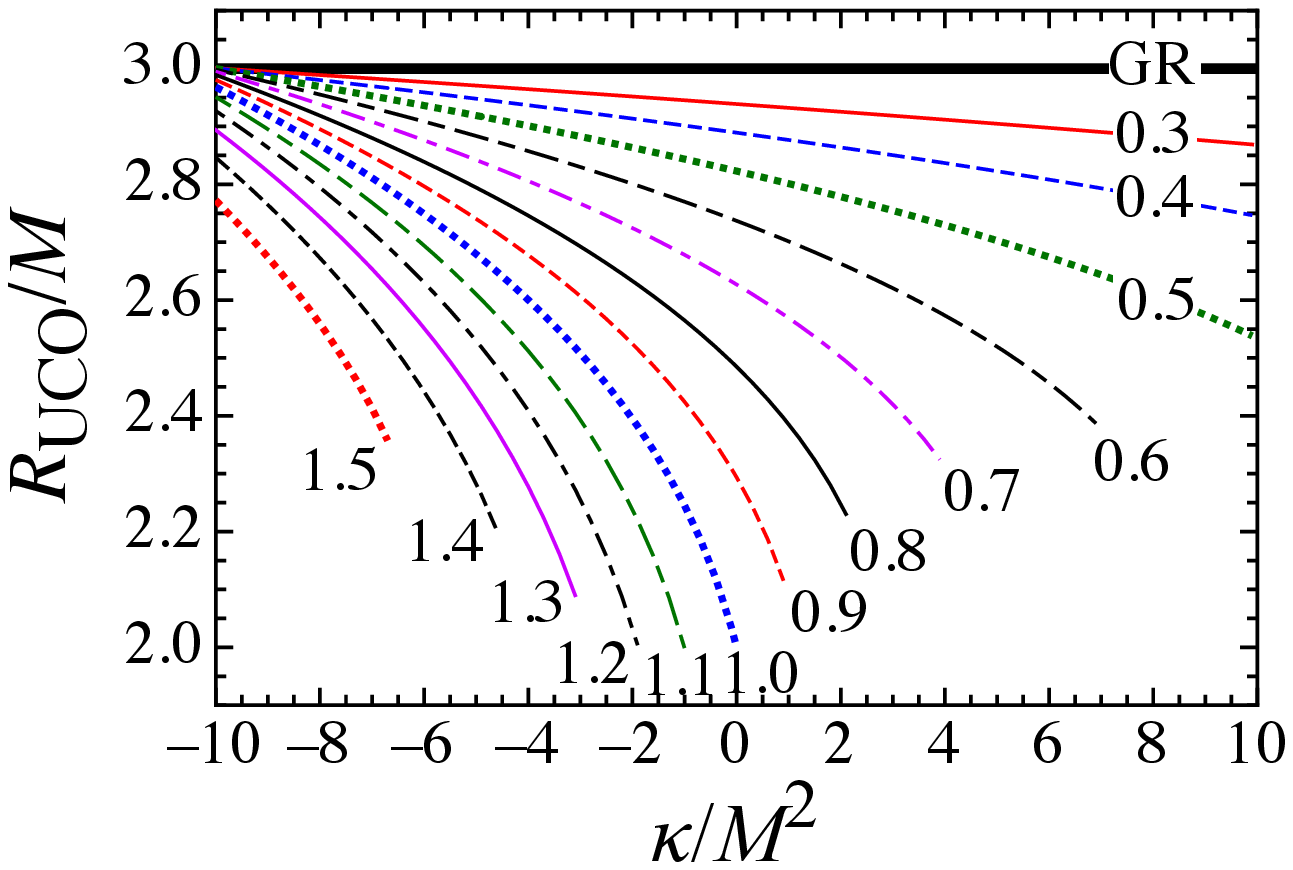} 
\end{tabular}
\end{center}
\caption{
Radius of the unstable circular orbit $R_{\rm UCO}$ around the electrically charged black hole in EIBI as a function of the electrical charge $Q/M$ with the fixed value of the coupling constant $\kappa/M^2$ (left panel) and as a function of $\kappa/M^2$ with fixed value of $Q/M$ (right panel). In the figure, the thick-solid line corresponds to the result in general relativity, i.e., for the Reissner-Nordstr\"om spacetime in the left panel and for the Schwarzschild spacetime in the right panel.
}
\label{fig:RUCO}
\end{figure*}

\begin{figure*}
\begin{center}
\begin{tabular}{cc}
\includegraphics[scale=0.5]{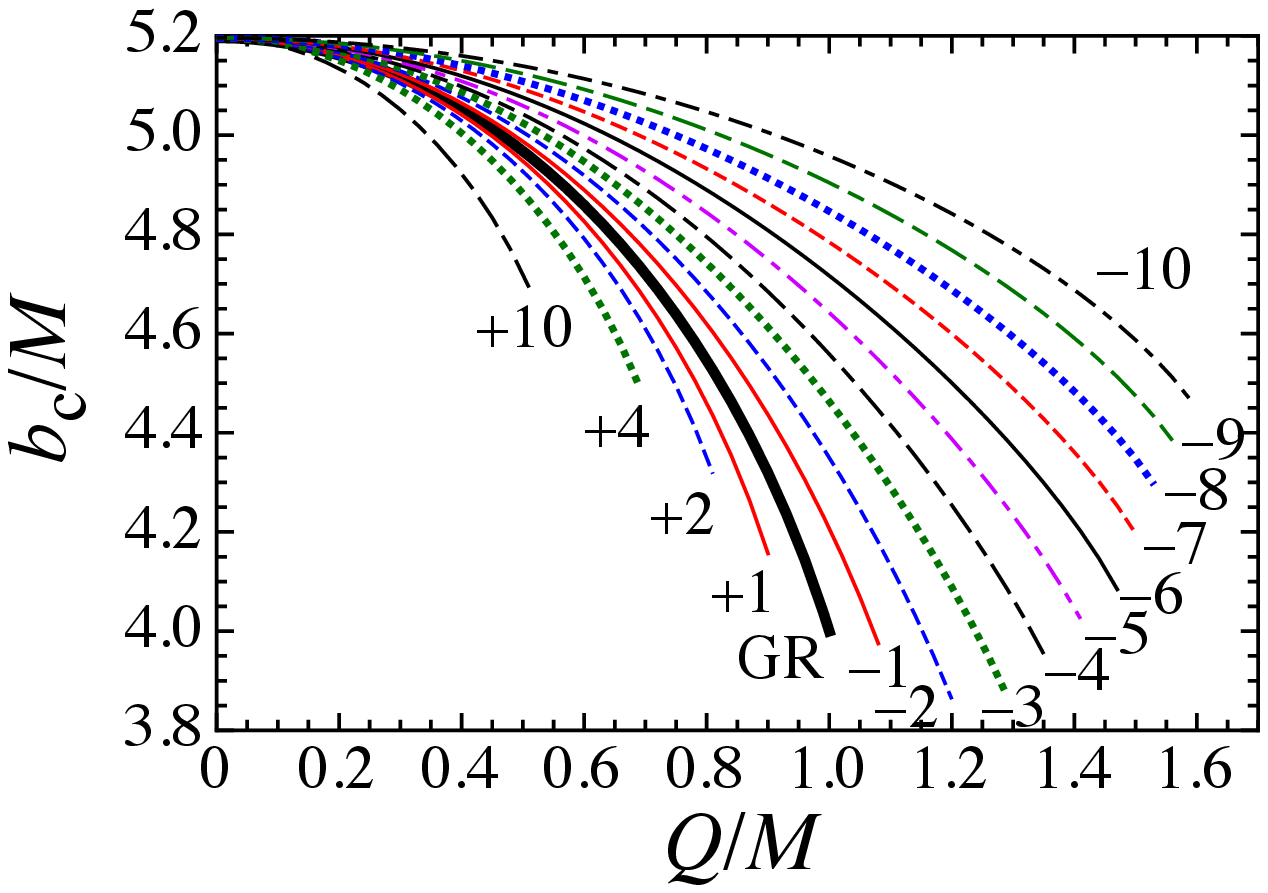} &
\includegraphics[scale=0.5]{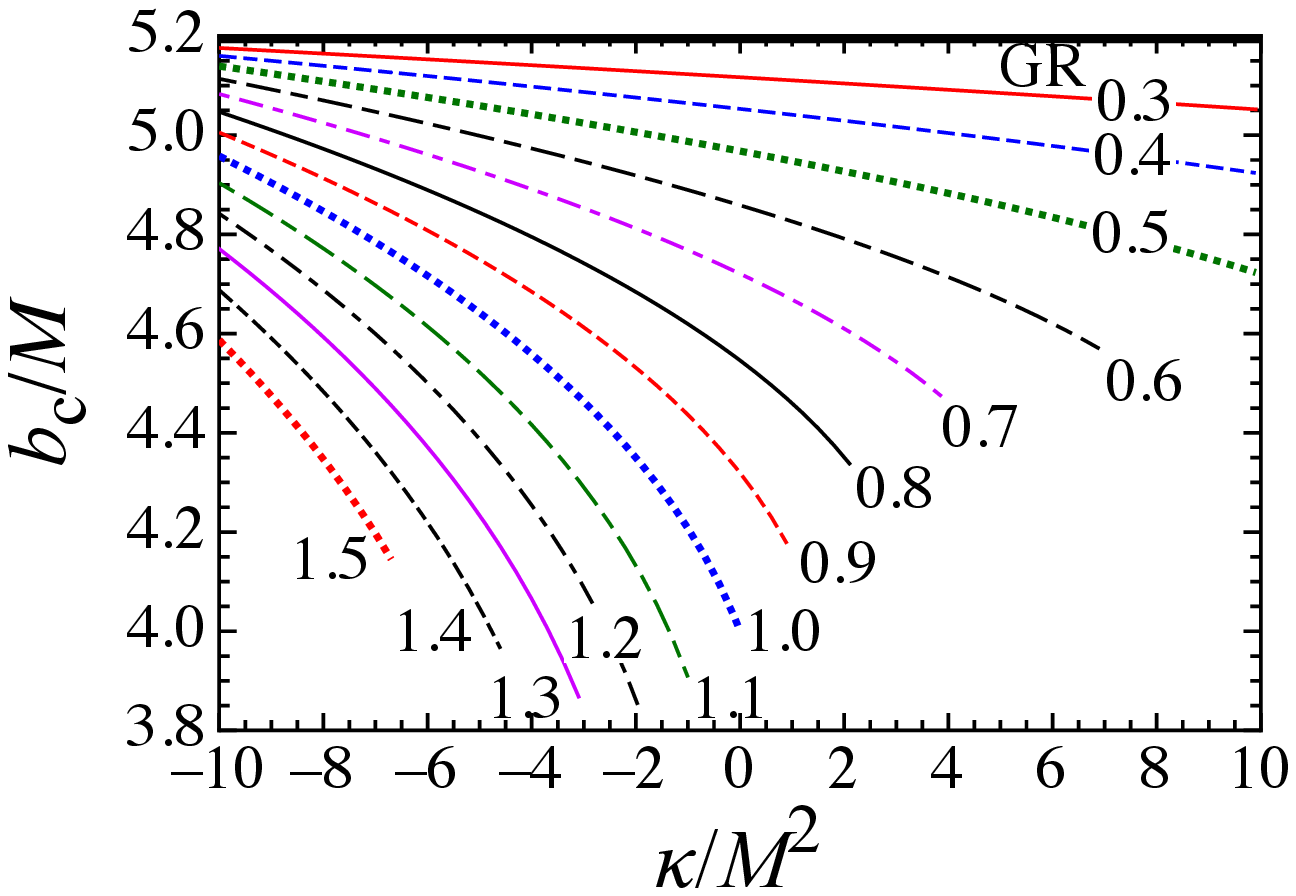} 
\end{tabular}
\end{center}
\caption{
Critical impact parameter $b_c$ in EiBI as a function of the electrical charge $Q/M$ with the fixed value of the coupling constant $\kappa/M^2$ (left panel) and as a function of $\kappa/M^2$ with fixed value of $Q/M$ (right panel). As same as in Fig.\ref{fig:RUCO}, the thick-solid line corresponds to the result in general relativity.
}
\label{fig:bc}
\end{figure*}

In Fig. \ref{fig:RUCO}, we show the dependence of $R_{\rm UCO}$ in EiBI on the electrical charge $Q/M$ with fixed value of the coupling constant $\kappa/M^2$ in the left panel and on $\kappa/M^2$ with fixed value of $Q/M$ in the right panel, where the labels in the figure denote the fixed value of $\kappa/M^2$ in the left panel and the fixed value of $Q/M$ in the right panel. For comparison, we also show $R_{\rm UCO}$ in general relativity with the thick-solid line, i.e.,  for the Reissner-Nordstr\"om spacetime in the left panel and for the Schwarzschild spacetime in the right panel. From this figure, one can observe that $R_{\rm UCO}$ depends strongly not only on the value of $Q/M$ but also on the value of $\kappa/M^2$. In particular, 
$R_{\rm UCO}$ with a fixed value of $Q$ is a monotonically decreasing function of the coupling constant $\kappa$. Note that the unstable circular orbit exists, just like the event horizon \cite{SM2014}, even for $Q/M>1$, provided $\kappa < 0$. On the other hand, the critical impact parameter $b_c$ with which the massless particle moves on the UCO, also depends on the values of $Q/M$ and $\kappa/M^2$. In Fig. \ref{fig:bc}, we show the dependence of $b_c$ on $Q/M$ in the left panel and on $\kappa/M^2$ in the right panel, as same as in Fig. \ref{fig:RUCO}. From this figure, we find that the dependence of $b_c$ on $Q/M$ and $\kappa/M^2$ are qualitatively the same as that of $R_{\rm UCO}$. We also find that $b_c$ in EiBI cannot be over $b_c$ for the Schwarzschild spacetime, which is $b_c=3\sqrt{3}M$, in the range of the parameters we adopt in this paper.

\section{Deflection angle}
\label{sec:IV}

\begin{figure*}
\begin{center}
\begin{tabular}{cc}
\includegraphics[scale=0.4]{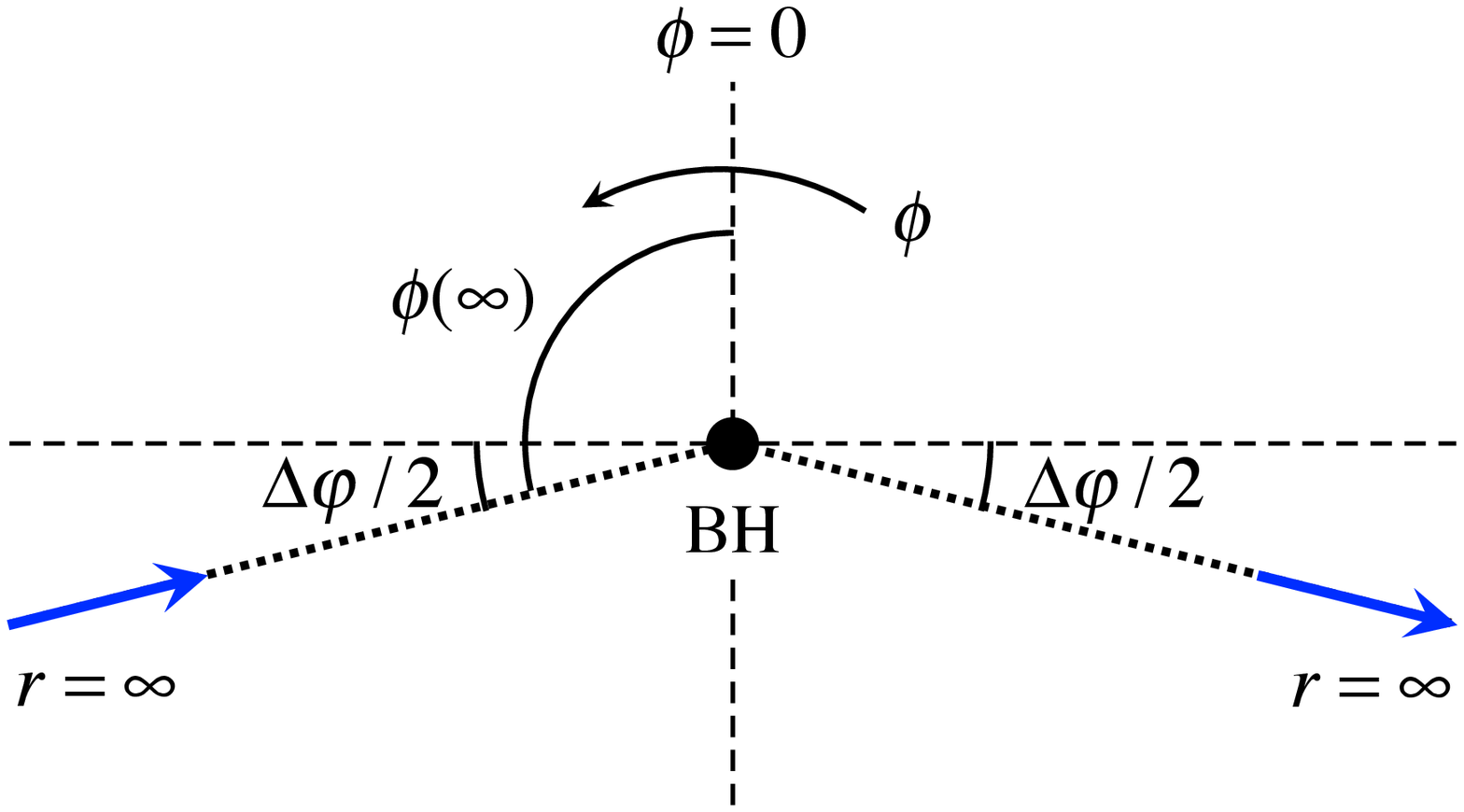} &
\includegraphics[scale=0.4]{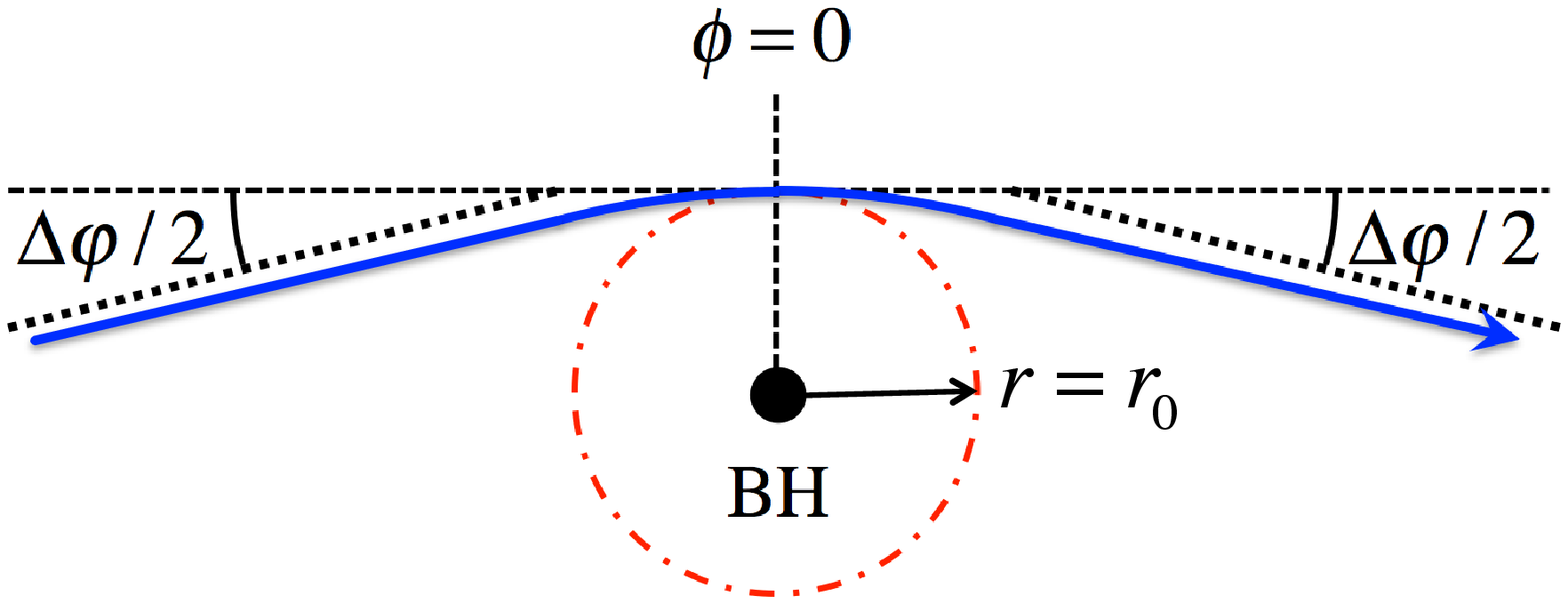}
\end{tabular}
\end{center}
\caption{
Image of scattering orbit of massless particle, where $\Delta\varphi$ is deflection angle. The right panel is magnified view in the vicinity of the scattering position, where $r_0$ is the curtate distance between the particle orbit and the black hole. 
}
\label{fig:dphi}
\end{figure*}

\begin{figure*}
\begin{center}
\begin{tabular}{cc}
\includegraphics[scale=0.5]{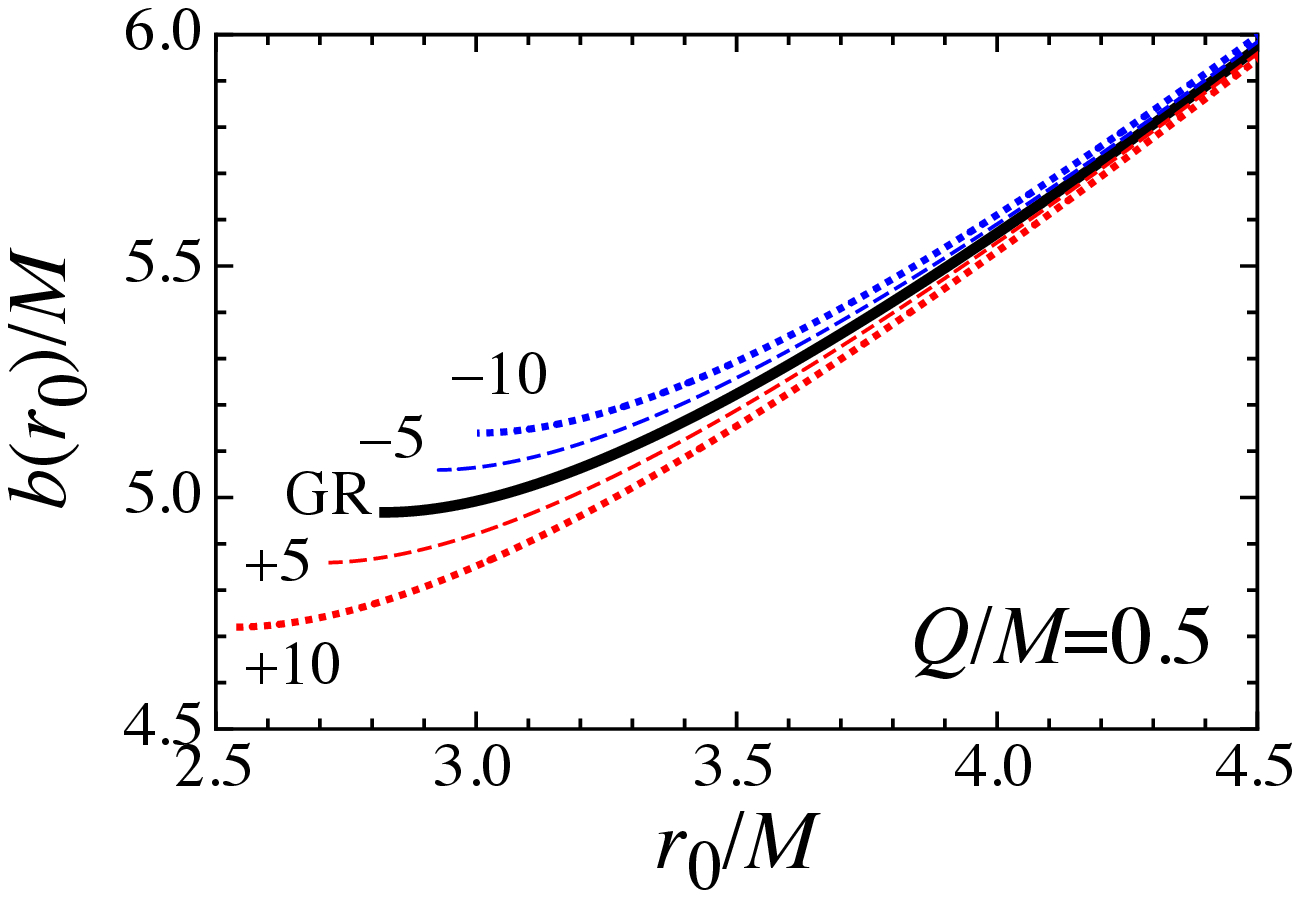} & 
\includegraphics[scale=0.5]{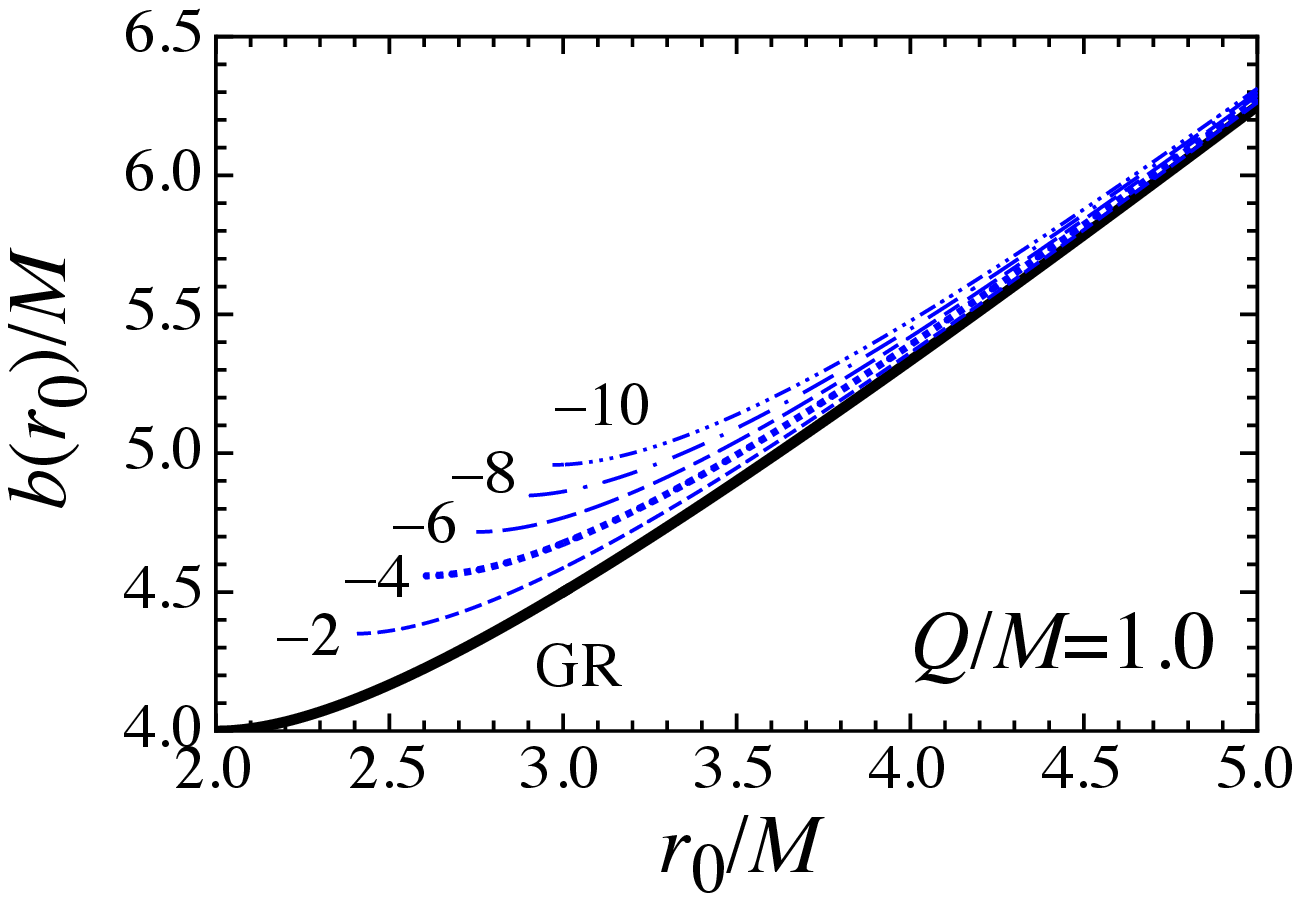} 
\end{tabular}
\end{center}
\caption{
For the scattering massless particle, impact parameter in EiBI is shown as a function of the position of the turn-around point, where the left and right panels correspond to the results for $Q/M=0.5$ and $1.0$, respectively. The labels in the figure denote the value of the coupling constant $\kappa/M^2$. For reference, the case in general relativity ($\kappa=0$) is also shown with the thick-solid line.}
\label{fig:br0}
\end{figure*}

We especially consider the scattering orbit of massless particle around the electrically charged black hole in EiBI, i.e, for $b>b_c$. Assuming the particle motion on the equatorial plane and using Eqs. (\ref{eq:tphi}) and (\ref{eq:geodesic}), one can derive the equation,
\begin{equation}
  \frac{d\phi}{dr} = \frac{d\phi/d\tau}{dr/d\tau} = \frac{b\psi}{r\sqrt{r^2 - f\psi^2b^2}}. \label{eq:dphi_dr}
\end{equation}
As shown in Fig. \ref{fig:dphi}, we consider that the particle turns around the black hole at $r=r_0$ and $\phi=0$. We remark that the position of $r_0$ in the effective potential (Fig. \ref{fig:Vr}) corresponds to this turn-around point. Since, by definition, $dr/d\phi$ should be zero at the turn-around point, the impact parameter for $b\ge b_c$ can be expressed as
\begin{equation}
  b^2 = \frac{r_0^2}{f(r_0)\psi^2(r_0)}. \label{eq:br0}
\end{equation}
In Fig. \ref{fig:br0}, the impact parameter determined by Eq. (\ref{eq:br0}) is shown as a function of the position of the turn-around point $r_0$, where the left and right panels correspond to the results for $Q/M=0.5$ and $1.0$, respectively. In the figure, the labels denote the values of $\kappa/M^2$, where the case in general relativity ($\kappa=0$) is also plotted by the thick-solid line for reference. The leftmost end-point of each curve corresponds to the critical impact parameter $b_c$ for each case, which is given by $b_c^2=R_{\rm UCO}^2/[f(R_{\rm UCO}) \psi^2(R_{\rm UCO})]$. From this figure, we find that the impact parameter in EiBI for the massless particle scattering far from the black hole, such as $r_0\gtrsim 5M$, is almost independent from the coupling constant, while that depends strongly on the coupling constant and the electrical charge for the scattering orbit close to the black hole.

Additionally, integrating Eq. (\ref{eq:dphi_dr}), we can get the equation,
\begin{equation}
 \phi(r) - \phi(r_0) = \int_{r_0}^r \frac{b\psi}{r\sqrt{r^2 - f\psi^2b^2}}dr.
\end{equation}
Therefore, as shown in Fig. \ref{fig:dphi}, the deflection angle $\Delta \varphi(r_0)$ is determined as
\begin{eqnarray}
	\Delta \varphi &=& 2\phi(\infty) - \pi \nonumber \\
	  &=& 2b\int_{r_0}^\infty \frac{\psi}{r\sqrt{r^2 - f\psi^2b^2}}dr - \pi, \label{eq:dphi}
\end{eqnarray}
where we adopt $\phi(r_0)=0$. However, one can see that the integrand in Eq. (\ref{eq:dphi}) diverges at $r=r_0$ due to Eq. (\ref{eq:br0}). That is, as shown in \cite{B2002,Wei:2014dka}, one should remove the pole at $r=r_0$ for determining the deflection angle $\Delta \varphi$.

Now, introducing a new variable $z$ defined as $z\equiv 1-r_0/r$, i.e., $r(z)=r_0/(1-z)$, the angle $\phi(\infty)$ can be express as
\begin{equation}
  {\cal I}(r_0) \equiv \phi(\infty) = \int_0^1 \psi (z) {\cal F}(z,r_0)dz,
\end{equation}
where 
${\cal F}$ is given by
\begin{gather}
  {\cal F}(z,r_0) = \frac{br(z)}{r_0\sqrt{r^2(z) - f(z)\psi^2(z)b^2}}.
\end{gather}
As mentioned the above, the function of ${\cal F}(z,r_0)$ diverges at $z=0$ ($r=r_0$), while $\psi (z)$ is regular for any values of $z$. To see the behavior of divergence of the function ${\cal F}$ at $z=0$, we approximate it near $z=0$ as
\begin{equation}
	{\cal F}(z,r_0) \simeq \frac{1}{\sqrt{\alpha z + \beta z^2}} \equiv {\cal F}_0(z,r_0),
\end{equation}
where $\alpha$ and $\beta$ are functions of $r_0$ given as
\begin{eqnarray}
  \alpha(r_0) &=& \left[2f\psi^2 - \frac{d}{dz}(f\psi^2)\right]_{z=0} \nonumber \\
    &=& 2f_0\psi_0^2 - r_0(f_0\psi_0^2)', \\
  \beta(r_0) &=& \left[-f\psi^2 + 2\frac{d}{dz}(f\psi^2) - \frac{1}{2}\frac{d^2}{dz^2}(f\psi^2)\right]_{z=0} \nonumber \\
  &=& -f_0\psi_0^2 + r_0(f_0\psi_0^2)' - \frac{r_0^2}{2}(f_0\psi_0^2)''.
\end{eqnarray}
Here, the variables with the subscript $0$ denotes the corresponding values at $r=r_0$ or $z=0$, and the prime denotes the derivative with respect to $r_0$. 
Thus, one can find that the value of ${\cal I}(r_0)$ is finite when $\alpha$ is nonzero. Meanwhile, as mentioned in the previous section, $dV/dr=0$ is realized only at $r=R_{\rm UCO}$, which leads to the statement that $\alpha=0$ only at $r_0=R_{\rm UCO}$. So, the deflection angle is finite value for $r_0>R_{\rm UCO}$ and diverges at $r_0=R_{\rm UCO}$.

In order to determine the deflection angle by removing the singular point, we separate ${\cal I}(r_0)$ into two parts as
\begin{equation}
  {\cal I}(r_0) = {\cal I}_D(r_0) + {\cal I}_R(r_0), \label{eq:II}
\end{equation}
where
\begin{gather}
  {\cal I}_D(r_0) = \int_0^1 \psi_0{\cal F}_0(z,r_0)dz, \\
  {\cal I}_R(r_0) = \int_0^1 {\cal G}(z,r_0)dz, \\
  {\cal G}(z,r_0) \equiv \psi(z){\cal F}(z,r_0) - \psi_0{\cal F}_0(z,r_0).
\end{gather}
The first term in Eq. (\ref{eq:II}) can be analytically integrated \cite{B2002}, which becomes
\begin{equation}
   {\cal I}_D(r_0) = \frac{2\psi_0}{\sqrt{\beta}}\log\frac{\sqrt{\beta} + \sqrt{\alpha + \beta}}{\sqrt{\alpha}}. \label{eq:ID}
\end{equation}
On the other hand, for $r_0 > R_{\rm UCO}$, the function ${\cal G}(z,r_0)$ can be expanded around $z=0$ as
\begin{eqnarray}
   {\cal G}(z,r_0) &=& \left[\psi(z) - \psi_0\right]{\cal F}_0(z,r_0) \nonumber \\
      &=& \frac{d \psi(z)}{d z}\bigg|_{z=0}\sqrt{\frac{z}{\alpha(r_0)}} + {\cal O}(z),
\end{eqnarray}
which leads to ${\cal G}(0,r_0)=0$, i.e., ${\cal I}_R(r_0)$ becomes a finite value. 
Finally, one can calculate the deflection angle for $r_0>R_{\rm UCO}$ via the relation,
\begin{equation}
   \Delta \varphi(r_0) = 2 {\cal I}_D(r_0) + 2{\cal I}_R(r_0) - \pi.   \label{eq:Dphi0}
\end{equation}

\begin{figure*}
\begin{center}
\begin{tabular}{cc}
\includegraphics[scale=0.5]{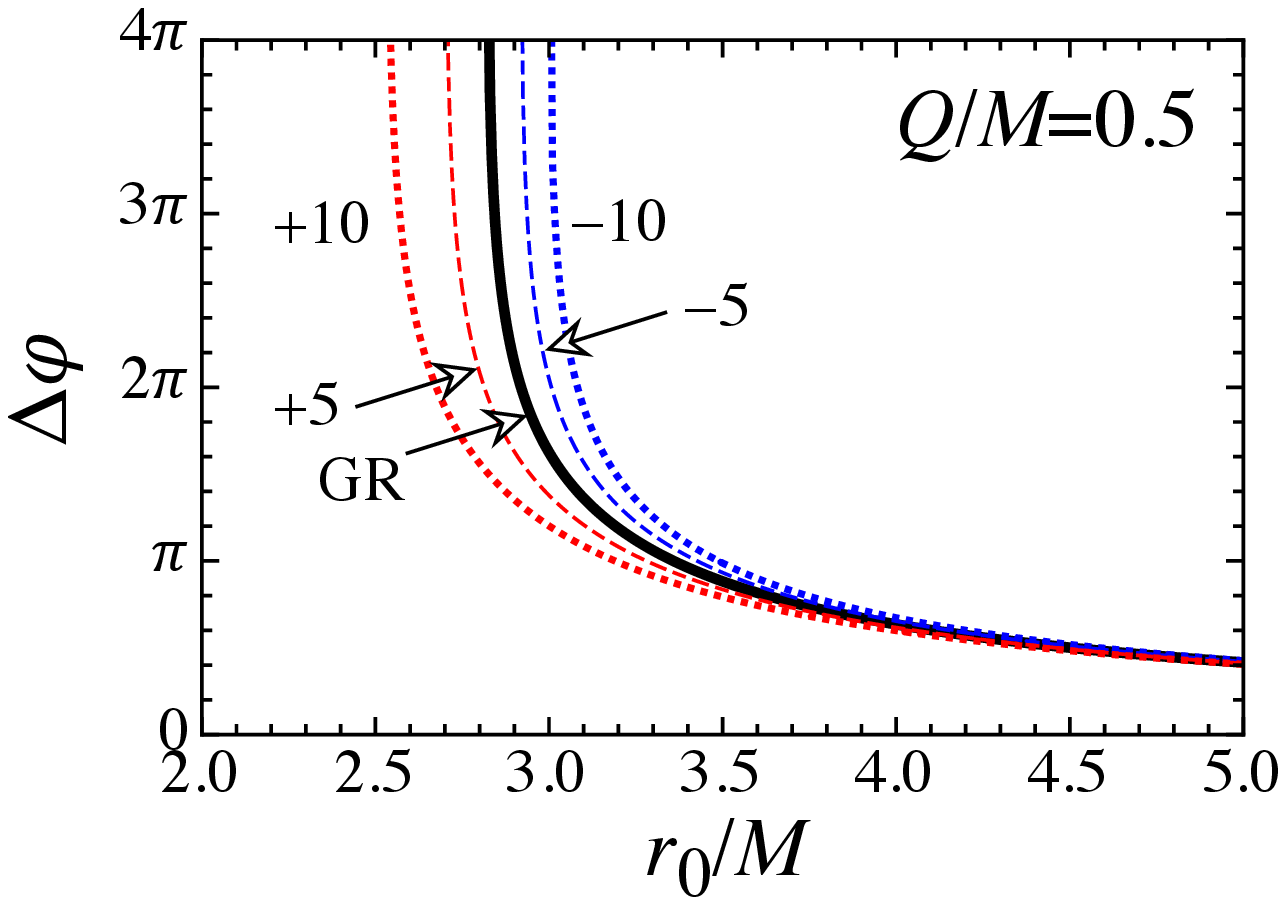} & 
\includegraphics[scale=0.5]{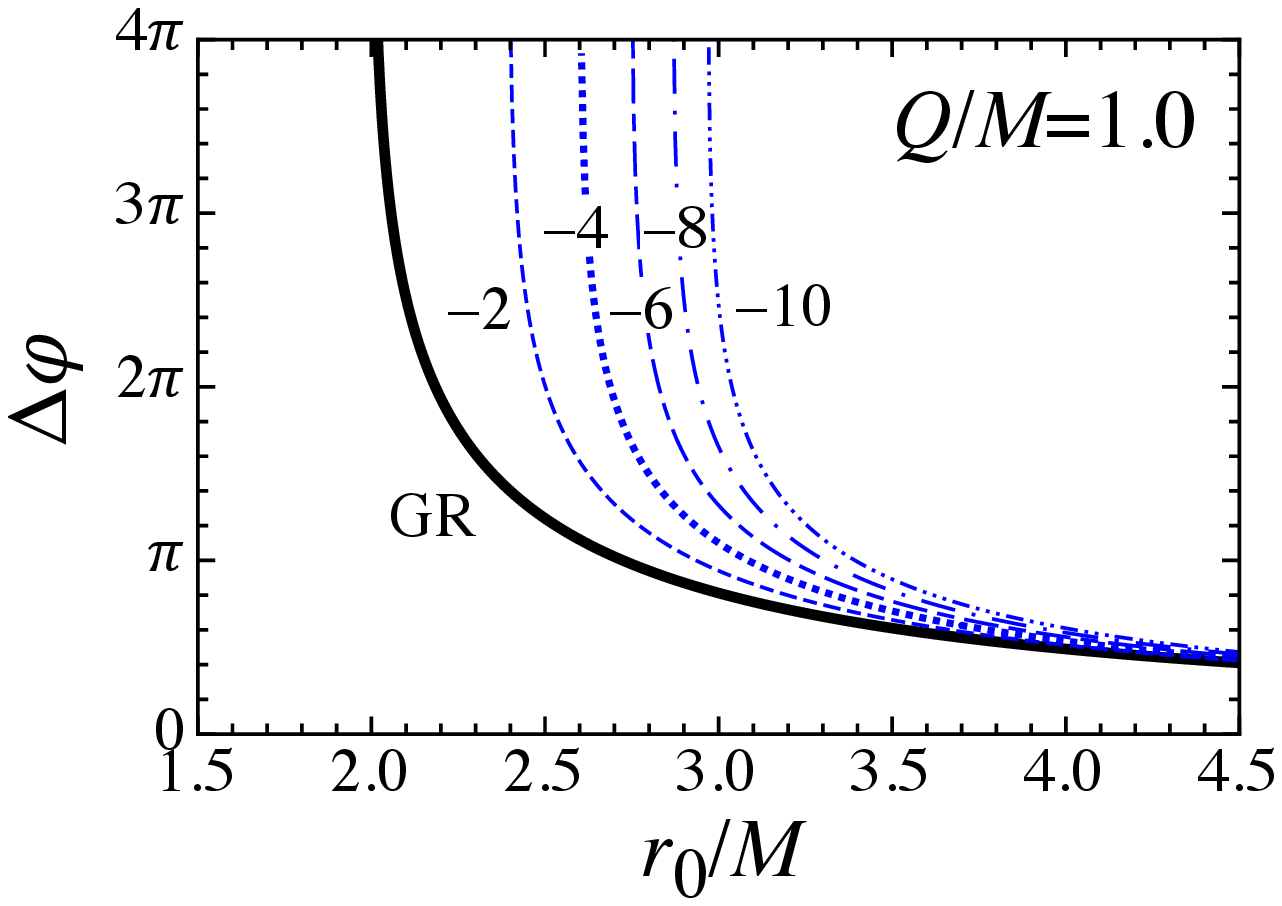} 
\end{tabular}
\end{center}
\caption{
Deflection angle is shown as a function of the position of the turn-around point. The left and right panels correspond to the results for $Q/M=0.5$ and $1.0$, respectively. The labels in the figure denote the value of the coupling constant $\kappa/M^2$ in EiBI, where the case in general relativity ($\kappa=0$) is also shown with the thick-solid line. 
}
\label{fig:Dphi-r0}
\end{figure*}

In Fig. \ref{fig:Dphi-r0}, we show the deflection angle as a function of $r_0$, where the left and right panels correspond to the results with $Q/M=0.5$ and 1.0, respectively. The labels in the figure denote the adopted values of the coupling constant $\kappa/M^2$ in EiBI. In any case, the deflection angle increases as $r_0$ decreases, and becomes equal to $2\pi$ with a specific value of $r_0$, where the massless particle goes around the lens object before reaching the observer. As $r_0$ additionally decreases, the deflection angle increases more and the number that the massless particle goes around the lens object also increases. When $r_0$ eventually reaches the position of the unstable circular orbit, the deflection angle diverges, where the massless particle goes around the lens object for all eternity. The positions of the relativistic images where the deflection angle becomes equal to a multiple of $2\pi$ depend strongly on the gravitational theory, as seen in Fig.\ref{fig:Dphi-r0}. In fact, the positions of the relativistic images for the negative (positive) coupling constant in EiBI become larger (smaller) than that expected in general relativity, where the difference between the positions expected in EiBI and in general relativity seems to increase with $Q/M$. On the other hand, since the deflection angle for $r_0\gtrsim 5M$ is almost independent from the gravitational theory, it could be quite difficult to distinguish the gravitational theory via the observation of the deflection angle for $r_0\gtrsim 5M$ even if detected.

\section{Strong gravitational lensing}
\label{sec:V}

Here, we focus on the deflection angle for $r_0 \simeq R_{\rm UCO}$, where $\alpha(r_0)$ is expanded as
\begin{gather}
  \alpha = \alpha_1(R_{\rm UCO})(r_0-R_{\rm UCO}) +{\cal O}((r_0-R_{\rm UCO})^2), \\
  \alpha_1(r_0) = (f_0\psi_0^2)' - r_0(f_0\psi_0^2)''.
\end{gather}
Substituting this expansion into Eq. (\ref{eq:ID}), one can get 
\begin{equation}
   {\cal I}_D(r_0) = -a_1\log\left(\frac{r_0}{R_{\rm UCO}} -1 \right) + \tilde{a}_2 + {\cal O}(r_0-R_{\rm UCO}),
\end{equation}
where $a_1$ and $\tilde{a}_2$ are the constants determined at $r_0=R_{\rm UCO}$ as 
\begin{gather}
  a_1 = \psi_0/\sqrt{\beta(R_{\rm UCO})}, \\
  \tilde{a}_2 = a_1\log\frac{4\beta(R_{\rm UCO})}{R_{\rm UCO}\alpha_1(R_{\rm UCO})}.
\end{gather}
The second term in Eq. (\ref{eq:II}) can be expanded in the vicinity of $r_0=R_{\rm UCO}$ as
\begin{eqnarray}
  {\cal I}_R(r_0) &=& \sum_{n=0}^\infty\frac{1}{n!}(r_0-R_{\rm UCO})^n\int_0^1\frac{\partial^n{\cal G}}{\partial r_0^n}\bigg|_{r_0=R_{\rm UCO}}dz 
       \nonumber \\
       &=& \int_0^1{\cal G}(z,R_{\rm UCO})dz + {\cal O}(r_0-R_{\rm UCO}).
\end{eqnarray}
Taking into account that $\alpha(R_{\rm UCO})=0$, ${\cal G}(z,R_{\rm UCO})$ can be expressed in the vicinity of $z=0$ as
\begin{eqnarray}
   {\cal G}(z,R_{\rm UCO}) &=& \left[\psi(z) - \psi_0\right]{\cal F}_0(z,R_{\rm UCO}) \nonumber \\
       &=& \frac{d \psi(z)}{d z}\bigg|_{z=0}\frac{1}{\sqrt{\beta(R_{\rm UCO})}} + {\cal O}(z).
\end{eqnarray}
That is, ${\cal I}_R(r_0)$ is still a finite value. Thus, one can calculate the deflection angle with
\begin{gather}
   \Delta \varphi(r_0) = -2a_1\log\left(\frac{r_0}{R_{\rm UCO}} -1 \right) + a_2 + {\cal O}(r_0-R_{\rm UCO}),  \label{eq:Dphi1} \\
  a_2 = 2\tilde{a}_2 +  2{\cal I}_R(R_{\rm UCO}) - \pi.
\end{gather}
Furthermore, from Eq. (\ref{eq:br0}), one can get the expansion of $b(r_0)$ around $r_0\simeq R_{\rm UCO}$ as
\begin{gather}
  b = b_c + \frac{1}{2}\frac{d^2b}{dr_0^2}\bigg|_{r_0=R_{\rm UCO}}\left(r_0-R_{\rm UCO}\right)^2.
\end{gather}
With this expansion, the deflection angle [Eq. (\ref{eq:Dphi1})] is rewritten as
\begin{gather}
  \Delta \varphi(b) =  -a_1\log\left(\frac{b}{b_c}-1\right) + a_3 + {\cal O}((b - b_c)^{1/2}), \label{eq:Dphi2} \\
  a_3 = a_2 + a_1\log\left(\frac{R_{\rm UCO}^2}{2b_c}\frac{d^2b}{dr_0^2}\bigg|_{r_0=R_{\rm UCO}}\right).
\end{gather}

\begin{figure}
\begin{center}
\includegraphics[scale=0.45]{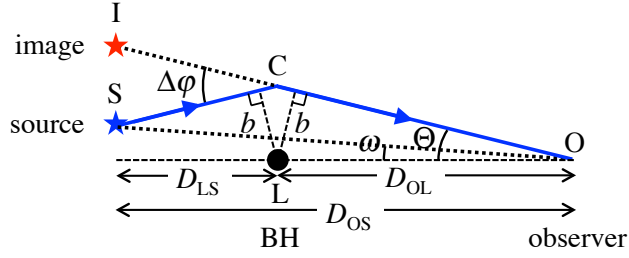} 
\end{center}
\caption{
Lens diagram, where S, L, O, and I are the positions of the source, lens object, observer, and image, respectively.
}
\label{fig:lens}
\end{figure}

Now, we consider the lens geometry shown in Fig. \ref{fig:lens}. We assume that the observer and source are located in the flat spacetime, while the region around the black hole is strongly curved spacetime, which leads to the light bending. That is, the light ray passes from the point S up to the point O via the point C in Fig. \ref{fig:lens}. Then, the lens equation can be expressed as \cite{VE2000}
\begin{equation}
  \tan\omega = \tan\Theta - \frac{D_{\rm LS}}{D_{\rm OS}}\left[\tan\Theta + \tan\left(\Delta\varphi-\Theta\right)\right],
\end{equation}
where $\omega$ corresponds to the angular separation between the lens and source, and $\Theta$ corresponds to that between the lens and image, i.e., $\omega =\angle$LOS and $\Theta =\angle$LOI in Fig. \ref{fig:lens}. One can also see that $b=D_{\rm OL}\sin\Theta$.

\begin{figure*}
\begin{center}
\begin{tabular}{cc}
\includegraphics[scale=0.5]{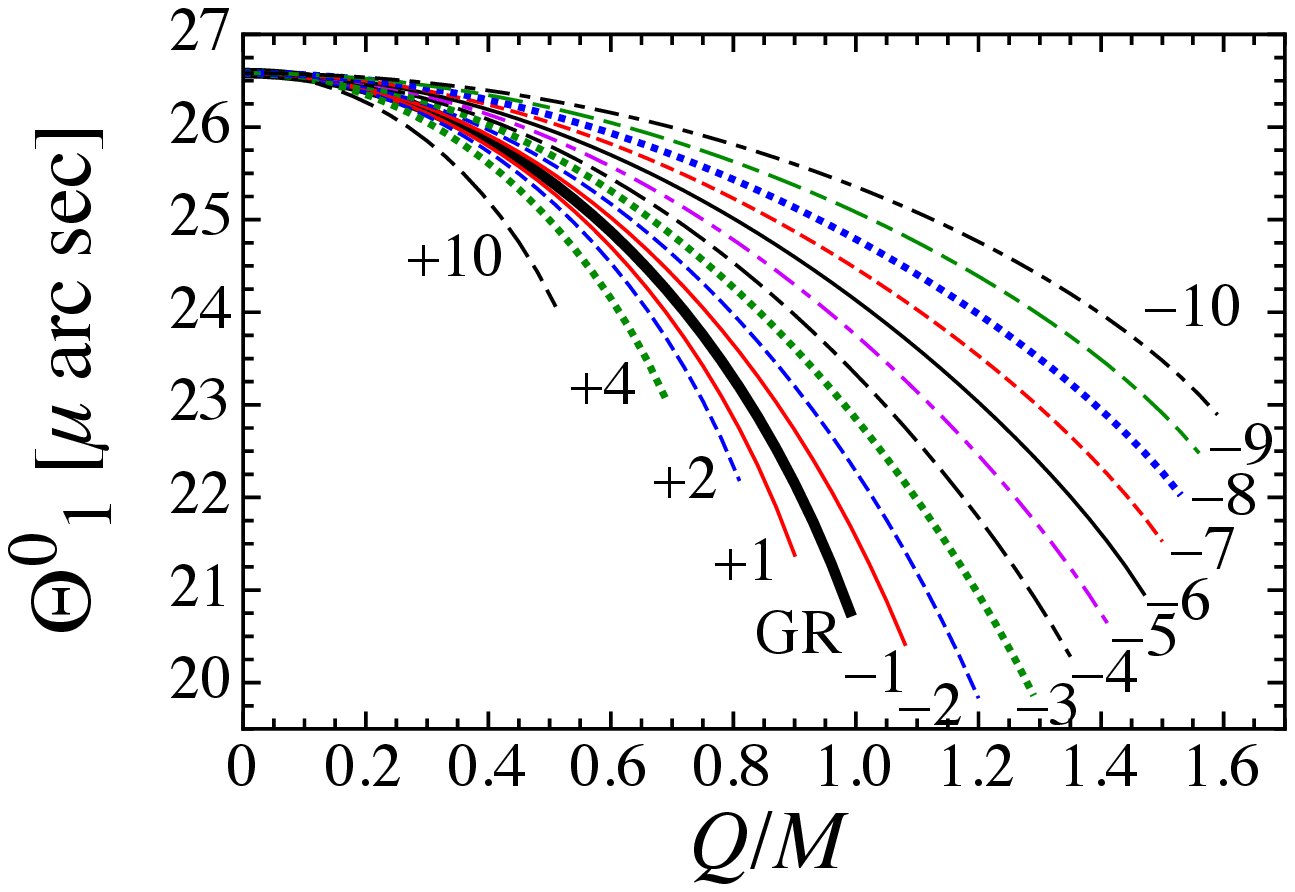} & 
\includegraphics[scale=0.5]{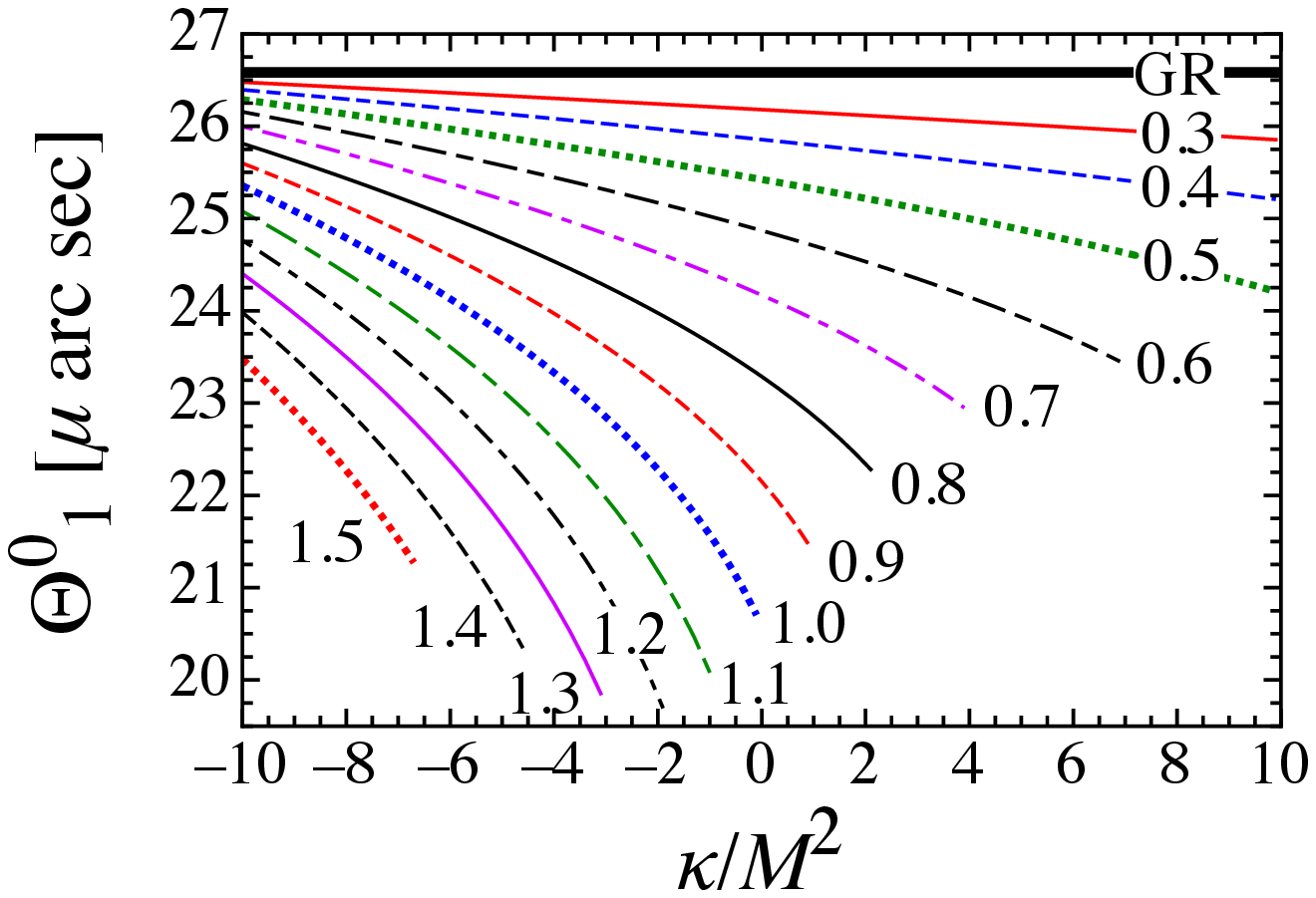} 
\end{tabular}
\end{center}
\caption{
The angle of the 1st relativistic image, $\Theta_1^0$, in EiBI as a function of the electrical charge $Q/M$ with the fixed value of the coupling constant $\kappa/M^2$ in the left panel and as a function of $\kappa/M^2$ with the fixed value of $Q/M$ in the right panel. The labels in the figure denote the fixed values of $Q/M$ or $\kappa/M^2$. For reference, the solid-thick line in each panel corresponds to the result in general relativity. 
}
\label{fig:T1}
\end{figure*}

In the strong deflection limit, i.e., $\omega\ll 1$, $\Theta\ll 1$, and $(\Delta\varphi_n-\Theta)\ll 1$, the lens equation reduces to
\begin{equation}
  \omega = \Theta - \frac{D_{\rm LS}}{D_{\rm OS}}\Delta\varphi_n,
\end{equation}
where $\Delta\varphi_n$ is the deflection angle removed all the loops of photon around the lens object \cite{B2002}. That is, $\Delta \varphi_n\equiv\Delta\varphi-2n\pi$ with an integer $n$ for $0<\Delta \varphi_n\ll 1$. In this limit, since $b\simeq D_{\rm OL}\Theta$, the deflection angle [Eq. (\ref{eq:Dphi2})] reduces to
\begin{equation}
  \Delta \varphi(\Theta) =  -a_1\log\left(\frac{D_{\rm OL}\Theta}{b_c}-1\right) + a_3. \label{eq:Dphi3}  
\end{equation}
Using Eq. (\ref{eq:Dphi3}), one can get the angle $\Theta_n^0$, with which $\Delta\varphi$ is equal to $2n\pi$, i.e.,
\begin{gather}
  \Theta_n^0 = \frac{b_c}{D_{\rm OL}}(1+e_n), \label{eq:Tn} \\
  e_n\equiv \exp\left(\frac{a_3-2n\pi}{a_1}\right).
\end{gather}
In Fig. \ref{fig:T1}, we show the angle of the 1st relativistic image, $\Theta_1^0$, determined from Eq. (\ref{eq:Tn}) as a function of the electrical charge $Q/M$ with the fixed value of the coupling constant in EiBI $\kappa/M^2$ in the left panel and as a function of the value of $\kappa/M^2$ with the fixed value of $Q/M$ in the right panel, where we assume that $D_{\rm OL}=8.5$ kpc and $M=4.4\times 10^6M_\odot$ to match the previous calculations in \cite{Wei:2014dka}. 
From this figure, we find that the angle of relativistic image for the fixed value of nonzero $Q/M$ decreases as the coupling constant increases.
In particular, the deviation from the result in general relativity significantly increases with the value of $|\kappa/M^2|$ and $Q/M$. This is consistent with the result in Fig. \ref{fig:Dphi-r0}, where 
the position of the turn-around point decreases as the coupling constant increases. 
In fact, compared with the results in general relativity, the value of $\Theta_1^0$ with $Q/M=0.5$ in EiBI becomes $3.4\%$ larger for $\kappa/M^2=-10$ and $4.8\%$ smaller for $\kappa/M^2=+10$, while $\Theta_1^0$ with $Q/M=1.0$ in EiBI becomes $22.5\%$ larger for $\kappa/M^2=-10$.

Setting $\Theta=\Theta_n^0+\Delta\Theta_n \; ( | \Delta\Theta_n / \Theta_n^0 | \ll 1)$ and expanding $\Delta\varphi(\Theta)$ around $\Theta_n^0$, one can derive $\Delta\varphi_n$ as a function of $\Delta\Theta_n$
\begin{equation}
  \Delta\varphi_n = -\frac{a_1 D_{\rm OL}}{b_ce_n}\Delta\Theta_n .
\end{equation}
Then, the lens equation becomes
\begin{eqnarray}
  \omega &=& \Theta_n^0 + \left(1 + \frac{a_1D_{\rm OL}}{b_ce_n}\frac{D_{\rm LS}}{D_{\rm OS}}\right)\Delta\Theta_n \nonumber \\
      &\simeq& \Theta_n^0 + \frac{a_1D_{\rm OL}}{b_ce_n}\frac{D_{\rm LS}}{D_{\rm OS}}\Delta\Theta_n,
\end{eqnarray}
where
we assume $D_{\rm OL}\gg b_c$ to obtain the final expression.
Then, the position of the $n$-th relativistic images, $\Theta=\Theta_n$, are given by
\begin{equation}
  \Theta_n = \Theta_n^0 + \frac{b_c e_n D_{\rm OS}}{a_1 D_{\rm LS} D_{\rm OL}}(\omega - \Theta_n^0).
\end{equation}
Additionally, the magnification of the images defined by the inverse of the Jacobian estimated at the images, $\mu_n$, is given by
\begin{eqnarray}
  \mu_n &\equiv& \left(\frac{\omega}{\Theta}\frac{\partial\omega}{\partial\Theta}\right)^{-1}\bigg|_{\Theta=\Theta_n^0} \nonumber \\
     &=& \frac{b_c e_n D_{\rm OS}}{a_1 D_{\rm OL} D_{\rm LS}},
\end{eqnarray}
where we simply consider the image at $\Theta=\Theta_n^0$ and we adopt the relation of $D_{\rm OL}\gg b_c$ again \cite{B2002}.

Finally, according to Ref. \cite{B2002}, we consider the simplest situation for observation. That is, we assume that only the outermost relativistic image $\Theta_1^0$ can be resolved from the others and all the remaining images are concentrated at $\Theta_\infty^0$. Then, the observable quantities are
\begin{gather}
  s \equiv \Theta_1^0 - \Theta_\infty^0 = \Theta_\infty^0 \exp(a_3/a_1 - 2\pi/a_1), \label{eq:ss} \\
  {\cal R} \equiv \mu_1\left(\sum_{n=2}^\infty\mu_n\right)^{-1} \simeq \exp(2\pi/a_1), \label{eq:RR}
\end{gather}
where we adopt the relations of $\exp(2\pi/a_1)\gg 1$ and $\exp(a_3/a_1)\sim{\cal O}(1)$ to derive the right hand side of Eq. (\ref{eq:RR}). We remark that the value of ${\cal R}$ is dependent on the properties of black hole spacetime, and independent of $D_{\rm OS}$, $D_{\rm LS}$, $D_{\rm OL}$, and $\omega$. We also remake that the values of $a_1$ and $a_3$ can be determined from the observable quantities $s$ and ${\cal R}$ as an inverse problem.

\begin{figure*}
\begin{center}
\begin{tabular}{cc}
\includegraphics[scale=0.5]{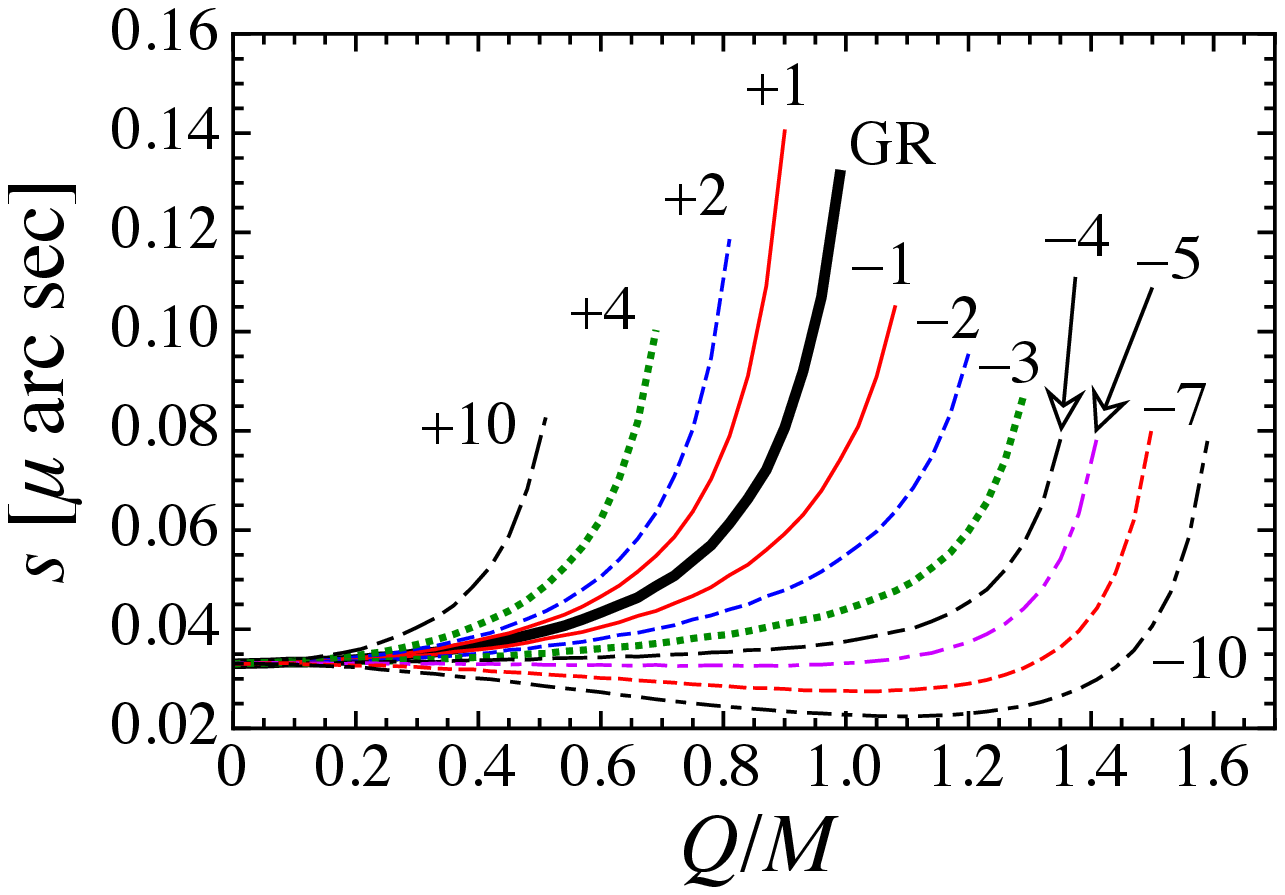} & 
\includegraphics[scale=0.5]{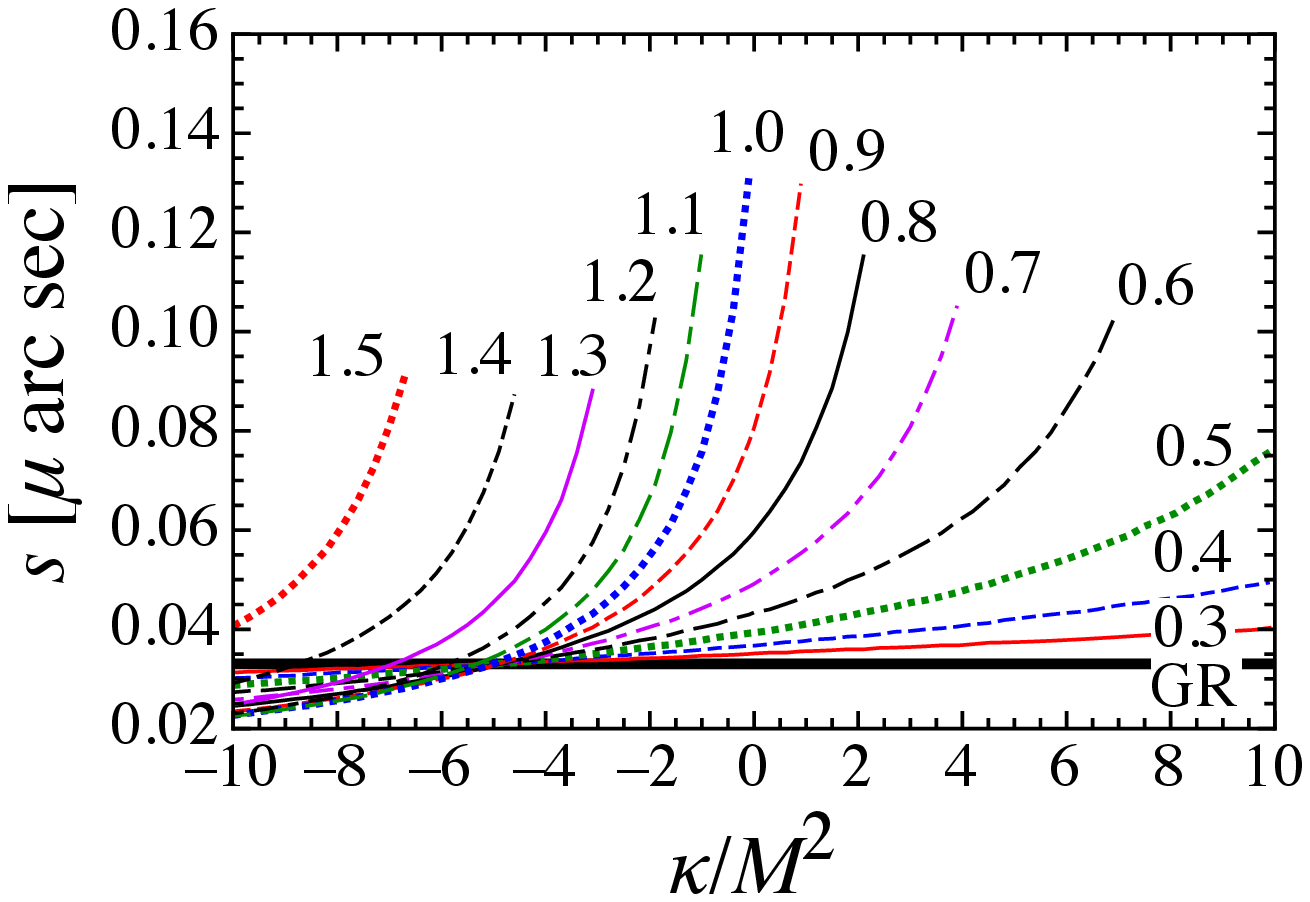} 
\end{tabular}
\end{center}
\caption{
Angular separation $s$ in EiBI as a function of the electric charge $Q/M$ with the fixed value of the coupling constant $\kappa/M^2$ in the left panel and as a function of the value of $\kappa/M^2$ with the fixed value of $Q/M$ in the right panel. 
}
\label{fig:ss}
\end{figure*}

\begin{figure*}
\begin{center}
\begin{tabular}{cc}
\includegraphics[scale=0.5]{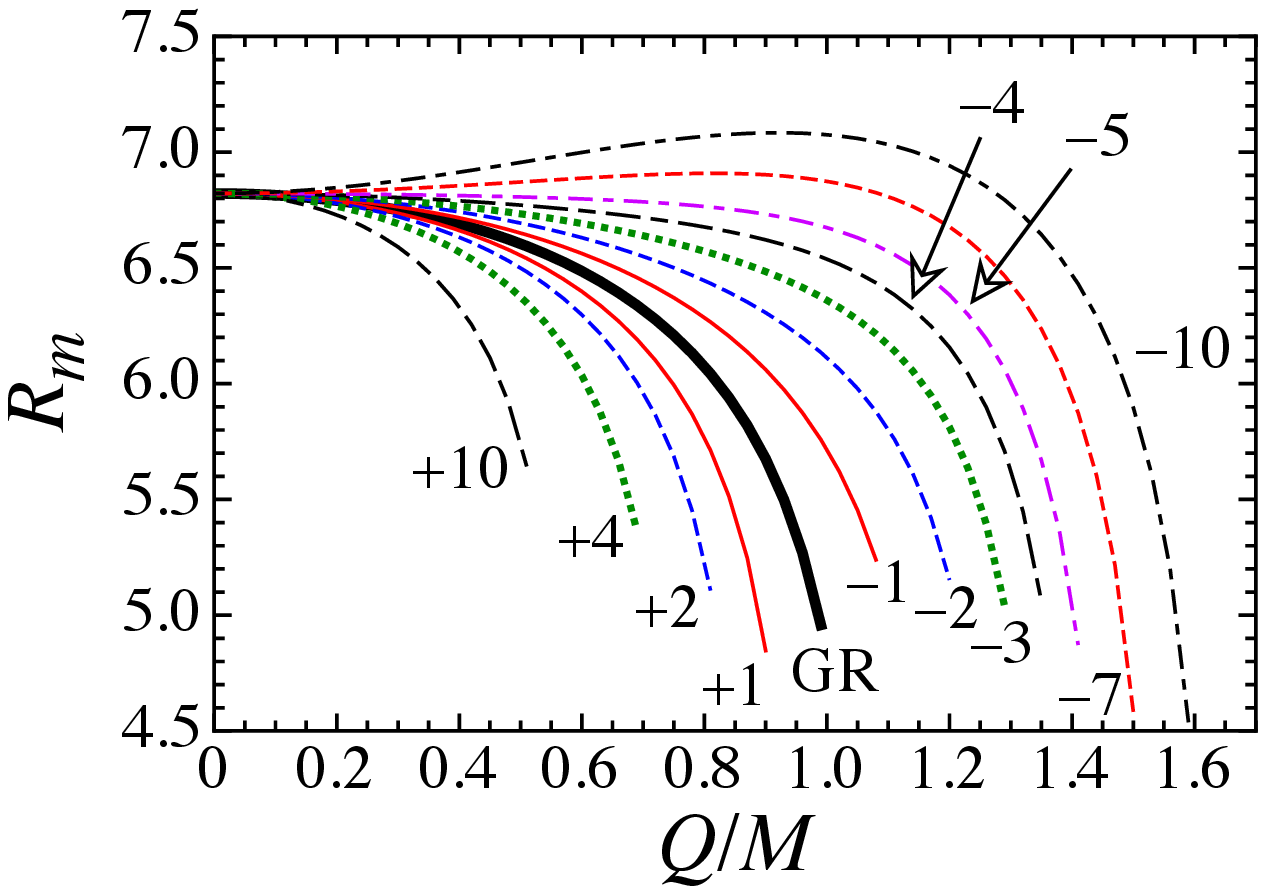} & 
\includegraphics[scale=0.5]{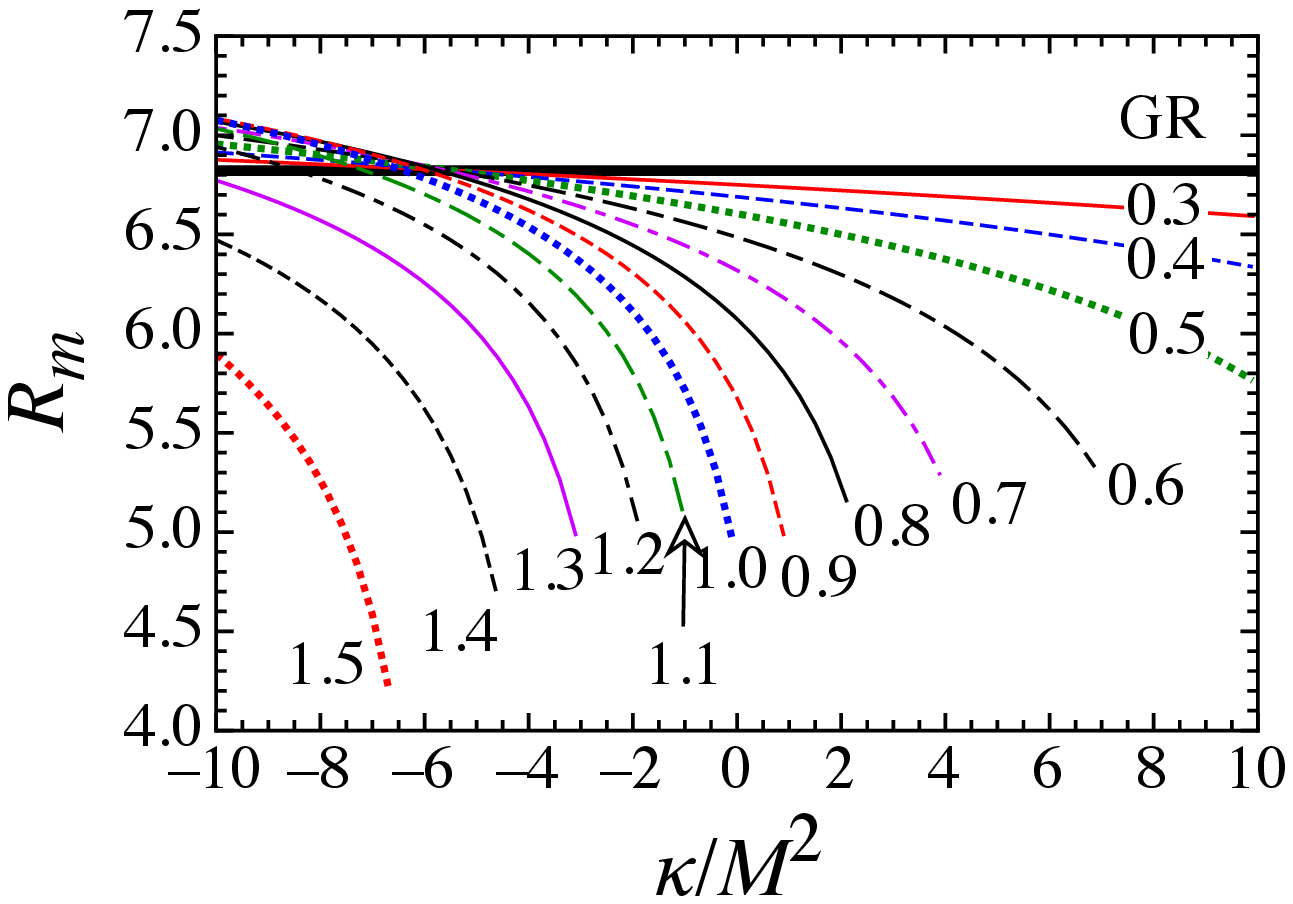} 
\end{tabular}
\end{center}
\caption{
Magnitude of the relative magnification ${\cal R}_m$ as a function of the electric charge $Q/M$ with the fixed value of the coupling constant $\kappa/M^2$ in the left panel and as a function of the value of $\kappa/M^2$ with the fixed value of $Q/M$ in the right panel. 
}
\label{fig:rm}
\end{figure*}

For example, we consider a specific case with the supermassive black hole located at the center of our Galaxy, whose mass is estimated to be $M=4.4\times 10^6M_\odot$ \cite{GEG2010}, while the distance from the solar system, $D_{\rm OL}$, is around 8.5 kpc \cite{note}. In Fig. \ref{fig:ss}, we show the angular separation between $\Theta_1^0$ and $\Theta_\infty^0$ as a function of the electric charge $Q/M$ with the fixed value of the coupling constant $\kappa/M^2$ in the left panel and as a function of the value of $\kappa/M^2$ with the fixed value of $Q/M$ in the right panel. From the right panel of this figure, one can observe that the angular separation $s$ with the fixed value of nonzero $Q/M$ increases with the coupling constant. In most cases, the separation angle in EiBI becomes larger than that for the Schwarzschild black hole, but $s$ can be smaller for $\kappa/M^2\lesssim -5$ and $Q/M\lesssim1.5$. Compared with the results in general relativity, i.e., for the Reissner-Nordstr\"om spacetime, the angular separation $s$ with $Q/M = 0.5$ in EiBI becomes $27.0\%$ smaller for $\kappa/M^2 = -10$ and $92.6\%$ larger for $\kappa/M^2 = +10$, while $s$ with $Q/M = 1.0$ in EiBI becomes $82.5\%$ smaller for $\kappa/M^2 = -10$, which could be a significant deviation from the expectation in general relativity. Moreover, in Fig. \ref{fig:rm}, the magnitude of the relative magnification ${\cal R}_m$, defined as
\begin{equation}
  {\cal R}_m \equiv 2.5\log_{10}{\cal R},
\end{equation}
is plotted as a function of the electric charge $Q/M$ with the fixed value of the coupling constant $\kappa/M^2$ in the left panel and as a function of the value of $\kappa/M^2$ with the fixed value of $Q/M$ in the right panel. 
We find that, with the fixed value of nonzero $Q/M$, the relative magnification decreases as the coupling constant increases in any case. 
In particular, in the case with $\kappa/M^2\lesssim -6$ and $Q/M\lesssim 1.3$, the relative magnification can be larger than that for the Schwarzschild black hole, which is the maximum value predicted in general relativity. That is, if the relative magnification larger than that expected for the Schwarzschild black hole would be observed, it could be the observation detecting an imprint of an alternative gravitational theory, not general relativity. Moreover, compared with the results in general relativity, the magnitude of the relative magnification with $Q/M = 0.5$ in EiBI becomes $5.5\%$ larger for $\kappa/M^2 = -10$ and $12.7\%$ smaller for $\kappa/M^2 = +10$, while ${\cal R}_m$ with $Q/M = 1.0$ in EiBI becomes $42.5\%$ larger for $\kappa/M^2 = -10$.

\section{Conclusion}
\label{sec:VI}

In this paper, we systematically examine the properties null geodesics around an electrically charged black hole in Eddington-inspired Born-Infeld gravity (EiBI), where we especially focus on the asymptotically flat black hole solution. In this case, the free parameters determining the back hole solution in EiBI are the coupling constant in EiBI ($\kappa/M^2$) and the electrical charge ($Q/M$), which are normalized by the mass of black hole $M$. In order to examine the orbital motion of a massless particle, we derive the null geodesic. We find that the radius of the unstable circular orbit, $R_{\rm UCO}$, depends strongly on both $\kappa/M^2$ and $Q/M$, where $R_{\rm UCO}$ with the fixed value of nonzero $Q/M$ decreases as $\kappa/M^2$ increases.

We also examine the strong gravitational lensing around the electrically charged black hole in EiBI. We show that the deflection angle due to the light bending in strong gravitational field diverges, when the distance between the black hole and the position of the turn-around point of light agrees with $R_{\rm UCO}$ as same as in general relativity. The position angles of the relativistic images also strongly depend on both $\kappa/M^2$ and $Q/M$, where the angle with the fixed value of nonzero $Q/M$ decreases with $\kappa/M^2$. Furthermore, in the strong deflection limit, we derive the analytic formulae to determine the angular position and magnification due to the winding of light around the black hole in EiBI, according to Ref. \cite{B2002}. In the simplest situation where only the outermost image can be resolved from the others, we concretely calculate the angular separation and the relative magnification, supposing the central black hole in our Galaxy. Then, we find that the angular separation and the relative magnification are quite sensitive to $\kappa/M^2$ and $Q/M$. In particular, the relative magnification in EiBI for $\kappa/M^2\lesssim -6$ and $Q/M\lesssim 1.3$ can be larger than the maximum value expected in general relativity. Since these observable quantities are directly affected by the parameters determining the black hole solution, it might be possible to make constraints on such parameters via the observation of the strong gravitational lensing \cite{BM2012}. Even if the current astronomical instruments are not enough to test the gravitational theory in a strong field regime, these observational constraints could enable us to reveal the gravitational theory in the future.

\acknowledgments
This work was supported in part by Grant-in-Aid for Young Scientists (B) through No. 26800133 (HS) provided by JSPS, by Grants-in-Aid for Scientific Research on Innovative Areas through No. 15H00843 (HS) provided by MEXT, and by Grant-in-Aid for Scientific Research (C) through No. 15K05086 (UM) provided by JSPS.



\end{document}